\documentclass[12pt]{article}
\usepackage{graphicx}
\usepackage{amsmath}
\usepackage{amssymb}
\usepackage{latexsym}
\usepackage{hyperref}
\usepackage[sort]{cite}

\newcommand{\be}{\begin{eqnarray}}
\newcommand{\ee}{\end{eqnarray}}
\newcommand{\bc}{\begin{center}}
\newcommand{\ec}{\end{center}}
\newcommand{\bea}{\begin{eqnarray}}
\newcommand{\eea}{\end{eqnarray}}
\newcommand{\ben}{\begin{equation}}

\newcommand{\nn}{\nonumber}

\numberwithin{equation}{section}

\newsavebox{\ns}
\newsavebox{\dbrane}
\newsavebox{\dbshort}

\usepackage{epsfig}
\usepackage{amssymb}

\def\appendix{{\newpage\section*{Appendix}}\let\appendix\section%
        {\setcounter{section}{0}
        \gdef\thesection{\Alph{section}}}\section}

\newcommand\ba{\begin{eqnarray}}
\newcommand\ea{\end{eqnarray}}

\def\Dslash{\,\,{\raise.15ex\hbox{/}\mkern-12mu D}}
\def\Dbarslash{\,\,{\raise.15ex\hbox{/}\mkern-12mu {\bar D}}}
\def\delslash{\,\,{\raise.15ex\hbox{/}\mkern-9mu \partial}}
\def\delbarslash{\,\,{\raise.15ex\hbox{/}\mkern-9mu {\bar\partial}}}
\def\pslash{\,\,{\raise.15ex\hbox{/}\mkern-9mu p}}
\def\calDslash{\,\,{\raise.15ex\hbox{/}\mkern-12mu {\cal D}}}

\newcommand{\hh}{{1\over 2}}

\renewcommand{\ll}{_}
\newcommand{\uu}{^}
\newcommand{\pp}{\partial}

\renewcommand{\exp}[1]{{\rm exp}\left( #1 \right)}

\renewcommand{\d}{\delta}
\newcommand{\m}{\mu}

\renewcommand{\m}{\mu}
\newcommand{\n}{\nu}
\newcommand{\s}{\sigma}
\renewcommand{\t}{\tau}
\newcommand{\G}{\Gamma}
\newcommand{\g}{\gamma}
\renewcommand{\a}{\alpha}
\renewcommand{\r}{\rho}

\renewcommand{\o}{\omega}
\newcommand{\e}{\epsilon}

\newcommand{\sqd}{^2}

\renewcommand{\hh}{{1\over 2}}

\newcommand{\eee}[1]{\ba{#1}\ea}
\renewcommand{\th}{\theta}
\renewcommand{\t}{\tau}

\renewcommand{\b}{\beta}

\newcommand{\pr}{^\prime {}}

\newcommand{\apr}{{\alpha^\prime} {}}

\newcommand{\IZ}{\relax\ifmmode\mathchoice
{\hbox{\cmss Z\kern-.4em Z}}{\hbox{\cmss Z\kern-.4em Z}}
{\lower.9pt\hbox{\cmsss Z\kern-.4em Z}} {\lower1.2pt\hbox{\cmsss
Z\kern-.4em Z}}\else{\cmss Z\kern-.4em Z}\fi} \font\cmss=cmss10
\font\cmsss=cmss10 at 7pt
\newcommand{\inbar}{\,\vrule height1.5ex width.4pt depth0pt}
\newcommand{\IC}{{\relax\hbox{$\inbar\kern-.3em{\rm C}$}}}
\newcommand{\IQ}{{\relax\hbox{$\inbar\kern-.3em{\rm Q}$}}}
\newcommand{\IP}{\relax{\rm I\kern-.18em P}}
\newcommand{\Ione}{{\relax\hbox{$\inbar\kern-.39em{\rm 1}$}}}

\newcommand{\ed}{\dot{e}}

\newcommand{\cc}{{\cal C}}

\newcommand{\ot}{\otimes}

\renewcommand{\cc}{{c_1}}

\renewcommand{\o}{\omega}

\newcommand{\ct}{\tilde{c}}

\renewcommand{\cc}{c}

\renewcommand{\pr}{{}^\prime{}}

\newcommand{\pst}{\tilde{\psi}}

\newcommand{\IR}{\relax{\rm I\kern-.18em R}}
\def\blfootnote{\xdef\@thefnmark{}\@footnotetext}

\renewcommand{\cc}[1]{\cite{#1}}

\newcommand{\bm}{\begin{matrix}}

\newcommand{\lba}{\left |}
\newcommand{\rba}{\right |}

\newcommand{\co}{{\cal O}}

\newcommand{\rr}[1]{(\ref{{#1}})}
\newcommand{\bbb}{\ba}
\renewcommand{\eee}{\ea}
\newcommand{\een}[1]{\label{#1}\ea}

\newcommand{\xxx}{(\xx)}

\newcommand{\prpr}{^{\prime\prime}{}}

\newcommand{\heading}[1]{\begin{center}\it {#1} \rm \end{center}}

\def\lrdd{\left(\,  }
\def\rrdd{\, \right)}
\def\lsqq{\left[ \,}
\def\rsqq{\, \right]}

\newcommand{\kket}[1]{\left | {#1} \right \rangle }

\def\bi{\begin{itemize}}
\def\ei{\end{itemize}}

\def\ed{\end{document}}

\def\cc{{\cal C}}

\renewcommand{\rr}[1]{(\ref{#1})}

\def\ct{{\cal T}}

\def\cc{\,}

\def\mfw{(-1)\uu{\rm F_{\rm w}}}

\newcommand{\nnno}[1]{  {{#1}\atop {#1}   }  }

\def\nno{\nnno{\circ}}

\def\xxx{\nn\\\nn\\}

\def\psq{^{\prime 2}}
\def\wsbb{worldsheet SUSY-breaking bubble~}
\def\tlc{{T^{\rm LC}}}
\def\gzpmbc{G^{\rm LC }}
\def\glc{G^{\rm LC}}
\def\cllc{{\cal L}_{\rm LC}}
\def\clc{c^{\rm LC}}
\def\ir{infrared~}
\def\uv{ultraviolet~}

\def\uvns{ultraviolet}
\def\zombie{ghost~}
\def\zombies{ghosts~}

\def\zombiesns{ghosts}

\def\mmm{\nn \\ && \nn \\}

\newcommand{\ssn}[1]{\nn\een{#1}}
\newcommand{\htil}{\tilde H}

\def\bfp{b_{\rm fp}}
\def\cfp{c_{\rm fp}}

\def\bone{{b_1}}
\def\cone{{c_1}}
\def\jbrst{J_{\rm BRST}}
\def\bos{{\rm mat}}
\def\quartet{{\rm quartet}}
\def\new{{\rm new}}
\def\deriv{{\rm deriv}}
\def\decup{{\rm decoupled}}
\def\tfif{{T}^{ (  {\rm mat} + b_1 c_1 ) }}
\def\bonecone{{b_1 c_1}}
\begin{document}

\begin{titlepage}
\begin{flushright}
hep-th/0612116
\end{flushright}
\vspace{15 mm}
\begin{center}
  {\Large \bf  Cosmological unification  \\ \vspace{.11in} of string theories  }  
\end{center}
\vspace{6 mm}
\begin{center}
{ Simeon Hellerman and Ian Swanson }\\
\vspace{6mm}
{\it School of Natural Sciences, Institute for Advanced Study\\
Princeton, NJ 08540, USA }
\end{center}
\vspace{6 mm}
\begin{center}
{\large Abstract}
\end{center}
\noindent
We present an exact solution of superstring theory
that interpolates in time between an initial type 0
phase and a final phase whose physics is exactly
that of the bosonic string.  
The initial theory is 
deformed by closed-string tachyon condensation along a lightlike direction.
In the limit of large tachyon vev, the worldsheet
conformal field theory precisely realizes
the Berkovits-Vafa embedding of bosonic string theory into superstring theory.  
Our solution therefore connects the bosonic string dynamically with the
superstring, settling a longstanding question about the relationship 
between the two theories.
\vspace{1cm}
\begin{flushleft}
December 13, 2006
\end{flushleft}
\end{titlepage}
\tableofcontents
\newpage

\section{Introduction}
We recently presented several exact classical solutions
of string theory in diverse dimensions, describing 
bulk closed-string tachyon condensation along a
lightlike direction  \cite{previous,previous2}.  
The solutions of \cite{previous} describe ``bubbles of nothing''
nucleating in unstable backgrounds, and Ref.~\cite{previous2} focuses on dynamical
transitions from unstable linear dilaton backgrounds to string theories in lower
numbers of dimensions.   In the latter case, the initial linear-dilaton theories
include bosonic, type 0 and unstable heterotic string theories.
They relax into final-state theories that can exist in a wide 
range of spacetime dimensions including critical, supercritical, and subcritical 
cases down to $2D$.
In all examples, however, the basic {\it kind} of string theory   
is unchanged between the initial and final configurations.  That is, 
the transitions in \cite{previous2} connect
bosonic string theories to other bosonic string theories, heterotic string theories
to one another, and type 0 theories to other type 0 or type II 
string theories, the two differing only in their GSO projection.

In this paper we study a related model of
lightlike tachyon condensation in type 0 string theory,
where the tachyon again depends only on
$X\uu +$, and is independent of the $D-2$
dimensions transverse to $X\uu \pm$.
In this example, the effect of the tachyon
condensate is not to change the number of
spacetime dimensions but to {\it change the kind of string theory altogether}.
Deep inside the region of nonzero tachyon condensate,
at $X\uu 0\to +\infty$, the string propagates freely
without a potential, but the worldsheet supersymmetry
is broken spontaneously.  The result is that
the effective worldsheet gauge symmetry and
constraint algebra are described by the ordinary
Virasoro algebra rather than its super-Virasoro extension,
and the physical state space is described by the
Hilbert space of a bosonic string rather than that
of a superstring.

It turns out that our solution precisely 
realizes an old mechanism \cite{bv} by
which bosonic string theory can be embedded
in the solution space of the superstring.
Since worldsheet supersymmetry is a gauge artifact, 
this mechanism is perfectly consistent.  
We note that the model in \cite{bv} is itself a 
specific instance of a very general idea, namely
the completely nonlinear realization
of supersymmetry \cite{va}.
The additional input of \cite{bv} is essentially 
conformal invariance and coupling to two-dimensional gravity.

The meaning
of the embedding of \cite{bv} has remained obscure.
One might have interpreted the bosonic string
as merely a formal  solution to the equations
of superstring theory, disconnected from the 
more familiar solution space.  
We are led to ask:  Is bosonic string theory
accessible dynamically from a conventional
state of superstring theory?

We settle this question in the affirmative.
In the case we study here, the embedding
of \cite{bv} describes only the $X\uu +\to +\infty$ 
limit of the theory (which we refer to as the \ir limit).   
The $X\uu +\to -\infty$ (or \uvns) limit
is described by a type 0 superstring, with
worldsheet supersymmetry linearly realized on free
fields in the conventional way.
At finite $X\uu +$, the system is a deformation interpolating between 
the two.  Our solution therefore represents a {\it dynamical transition}
between the type 0 superstring and the bosonic string.

Key to the analysis is the fact that the solutions we consider
have no dynamical quantum corrections.
This can be seen at the level of Feynman diagrams.
Our worldsheet theory is a theory of free
massless scalar fields, perturbed by an interaction
term depending only on $X\uu +$ and its superpartners
$\psi\uu +,~ \pst\uu +$.  Worldsheet propagators are
oriented, pointing from fields in the $X\uu +$ 
multiplet to the $X\uu -$ multiplet.  Since interaction
vertices have only outgoing lines, connected 
Feynman diagrams have at most one interaction vertex
(similar considerations were employed in \cite{previous,previous2}).

It follows
that connected correlators of fields in the $X^+ = -\infty$ limit 
have no loop corrections, and they can receive contributions from
at most one tree graph.  Even then, all operators in the
expectation value must be in the multiplet of $X\uu -$,
with no $X\uu +$ multiplet degrees of freedom.
By the same diagrammatic argument, connected interaction
contributions to
operator product expansions can have only $X\uu -$ multiplets
on the left-hand side and only $X\uu +$ multiplets on the
right-hand side.  We exploit these properties to understand
the large-$X\uu +$ behavior of the theory by
solving the OPE altogether.\footnote{To be precise, we solve the OPE
on a flat worldsheet.  We do not compute worldsheet curvature corrections to the
OPE, which could be relevant to the computation of scattering amplitudes.}

In the next section we discuss tachyon condensation in the
type 0 theory.  We find a solution, exact in $\apr$,
in which the tachyon 
increases exponentially along the lightlike direction $X\uu +$.
The $2D$ CFT has a large interaction term
in the limit $X\uu +\to +\infty$, so we choose a dual set of variables
whose interactions are suppressed, rather than enhanced, in the
limit of large $X\uu +$.  To gain an understanding of the basic physics,
we perform the change of variables classically, ignoring
singularities in products of local operators.
In Section 4 we work out
operator ordering prescriptions and derive the transformation on the
supercurrent and stress tensor at the quantum level.
In Section 5 we perform additional transformations
to make manifest the $(D-1)$-dimensional Poincar\'e invariance of
the $X\uu +\to +\infty$ limit.  Rewriting the stress tensor and supercurrent in an
$SO(D-1)$-symmetric form, we see that the large-$X\uu +$ limit of
our theory is exactly described by a $(1,1)$
superconformal field theory \cite{bv}, wherein the worldsheet supersymmetry is
completely nonlinearly realized.
In Section 6 we discuss the physics of the $X^+ \to +\infty$ limit,
and we present general conclusions in
Section 7.   Various technical issues are addressed in the appendices, 
including the presentation of specific canonical transformations, the 
direct computation of OPEs, the computation of the BRST cohomology in the
Berkovits-Vafa formalism, and a description of the Ramond sectors in this 
theory.

\section{Lightlike tachyon condensation in type 0 string theory}
We begin by considering type 0
strings propagating in a timelike linear dilaton
background in $D$ dimensions $X\uu M$, which has $G\ll{M N} = \eta\ll{M N}$
and $\Phi \equiv V\ll M X\uu M$, with $V\ll M$ constant.
This background describes a cosmological background
of type 0 string theory.  The cosmology is 
an expanding Friedman-Robertson-Walker solution
with flat spatial sections.  Its stress-energy comes
from the dynamics of the dilaton field $\Phi$,
whose potential gives rise to a quintessent cosmology
with equation of state
\bbb
w\equiv {p\over \rho} = - {{D - 3}\over{D - 1}}\ .
\eee
This value of $w$ is exactly on the boundary between
equations of state that give rise to accelerating and decelerating
behavior for the scale factor \cite{previous}.  Canonically
normalized fluctuations of the metric and dilaton have
exponentially growing modes, but these do not represent
true instabilities.  Their growth is always compensated by
the exponentially decreasing string coupling, so their coupling to
a mode of the string at most stays constant with time, and
in general decreases \cite{previous,evaofer}.  The same is true of
excited, massive modes of the string.  No massless or massive field
ever represents a true instability.

Only tachyons, the lowest modes of the string,
can grow fast enough with time to overcompensate
their decreasing coupling to the string worldsheet.
In \cite{previous,previous2} we examined certain
exact solutions in which the tachyon has a nonzero
value.  These solutions represent bubbles of
nothing, and dynamical transitions to string theories
with lower numbers of dimensions.  In this
paper we will examine a third type of exact solution
with nonzero tachyon.

Our starting point is the Lagrangian for a timelike linear dilaton
theory on a flat worldsheet, describing $D$ free massless fields
and their superpartners:
\bbb
{\cal L}\ll{\rm kin} = 
 {\frac{1}{2\pi}} G\ll{MN}
\lsqq
{\frac{2}{{\apr}}} 
 (\pp\ll + X\uu M)
(\pp\ll - X\uu N)
-  i \psi\uu M (\pp\ll - \psi\uu N) -  i \pst\uu M (\pp
\ll + \pst\uu N) \rsqq\ .
\een{type0freelag}
On a curved worldsheet, the dilaton couples to the worldsheet
Ricci scalar:
\bbb
\Delta {\cal L} = {1\over{4\pi}} \cc\Phi(X)~{\cal R}\ll{\rm worldsheet}\ .
\eee
The
magnitude of the dilaton gradient $V\ll{ M}$
must satisfy $4\apr V\sqd = -(D - 10)$, so we take
\bbb
V\ll + &=& V\ll - = - {{q}\over{\sqrt{2}}}\ ,
\mmm
\cc\cc V\ll i &=& 0,\qquad i =2,\cdots,D-1\ ,
\mmm
q &\equiv  &\sqrt{{D - 10}\over{4\apr}}\ ,
\een{vdef}
assuming the
dilaton rolls to weak coupling in the future.
The $\pm$ labels above refer to the light-cone directions 
$X\uu\pm\equiv {1\over{\sqrt{2}}} \lrdd X\uu 0 \pm X\uu 1 \rrdd$.
In addition to the massless modes of the metric $G\ll{MN}$,
NS B-field $B\ll{MN}$ and dilaton $\Phi$, there
is also a tachyon $\ct$ with mass-squared $m\sqd = - {2\over{\apr}}$.
Here and throughout the paper, we are using the terms 
{\it massive, massless, tachyonic, mass-squared,} etc.~in 
the sense described (for NS fields) in \cite{previous}.\footnote{Namely, when some NS 
field $\sigma$ enters the Lagrangian with mass term ${\cal L}\sim + e^{-2\Phi}m^2\sigma^2$,
we refer directly to $m^2$ as defining the mass.  Note that this is not the 
mass entering the dispersion relation of the canonically normalized field.  
See \cite{previous} for further discussion on this point.}

We would like to consider solutions for which
the type 0 tachyon condenses, growing exponentially in the
lightlike direction $X\uu +$.  That is, we make the ansatz 
\be
\ct \equiv  \tilde{\m} \cc \exp{\b X\uu +}\ .
\ee  
The linearized equation of motion for the tachyon is
\bbb
\pp\sqd \ct - 2 V\cdot \pp \ct + {2\over\apr} \ct = 0\ ,
\eee
which fixes
\bbb
\b q = {{\sqrt{2}}\over\apr}\ .
\een{betaqrel}
The type 0 tachyon
couples to the worldsheet as a $(1,1)$ superpotential:
\bbb
{\cal L}\ll{\rm int} = {i\over{2\pi}}\cc\int d\th\ll + d\th\ll -\cc
\ct (X)\ ,
\eee
which gives rise to a potential and Yukawa term
\bbb
{\cal L}\ll{\rm int} =
- {{\apr}\over{ 16 \pi}} G\uu{MN} \cc \pp\ll M\ct \pp\ll N\ct
+ {{i\apr}\over{4\pi}} \pp\ll M \pp\ll N \ct \cc\pst\uu M\psi\uu N \ ,
\eee
as well as a modified supersymmetry transformation for the fermions:
\bbb
\{Q\ll - , \psi\uu M\} &=& - \{Q\ll + ,\pst\uu M\} =  F\uu M \ ,
\mmm
F\uu M & \equiv & - \cc 
\sqrt{{\apr}\over 8}\cc
G\uu{MN} \pp\ll N{\ct} \ .
\eee
Since the gradient of the tachyon is null, the
worldsheet potential
$ {{\apr}\over{16\pi}} G\uu{MN} \cc \pp\ll M\ct \pp\ll N\ct$
is identically zero.  There is nonetheless a
nonvanishing $F$-term and a Yukawa coupling
between the lightlike fermions:
\bbb
F\uu - &=& + {{q\sqrt{\apr}\m }\over 2}\cc\exp{\b X\uu +}\ ,
\mmm
{\cal L}_{\rm Yukawa} &=& {{i\cc\m}\over{4\pi}}  \cc \exp{\b X\uu +}
\pst\uu +\psi\uu +  \ ,
\ee 
where $\m \equiv \b\sqd\apr\cc\tilde{\m}$.

Our goal is to determine the $X\uu + \to\infty$ limit of 
this theory.  Since there is no worldsheet potential, 
in contrast to the cases studied in \cite{previous,previous2}, 
no string states are expelled from the interior of 
the bubble.
Instead, all states are permitted within the
bubble: To describe the late-time behavior of this system
we must understand the dynamics of states in the $\ct\neq 0$ region.

The physics of the $X\uu +\to \infty$ phase 
may seem counterintuitive.  The nonvanishing $F$-term
signals spontaneously broken worldsheet supersymmetry,
but the vacuum energy vanishes.  This is not a
contradiction, as the worldsheet supersymmetry is realized
nonunitarily when treated as a global symmetry.  
If the supersymmetry is thought of as local, there is again no contradiction
because there are no generators of the superalgebra that act on the
physical-state Hilbert space.  From the latter point
of view, the vanishing of the vacuum energy in a theory
with spontaneously broken local supersymmetry is very much
like the physics of the so-called `no-scale' vacua
of $D = 4,~{\cal N} = 1$ supergravity.

This analogy lends a clue as to what the physics
in the $X\uu + \to + \infty$ limit might be, where the
$F$ term becomes infinite in size.  Consider
a no-scale vacuum of ${\cal N} = 1$ supersymmetry 
in four dimensions, in the limit where the $F$-term becomes large and the
vacuum energy is fine-tuned to zero.   If one
holds fixed the relevant parameters
of the theory other than the gravitino mass $m\ll{3/2}$,
the gravitino becomes infinitely massive and decouples.
In this limit one can introduce arbitrary small deformations
to the Lagrangian, including independent
variations of the masses of bose and fermi fields that would
otherwise lie in degenerate multiplets.
%
%
%
In other words, the system has a non-supersymmetric Minkowski vacuum 
that harbors no trace of the local supersymmetry of
the finite-$m\ll{3/2}$ theory.  

The $2D$ worldsheet CFT in our system behaves analogously:
the $X\uu +
\to \infty$ limit exhibits none of the local worldsheet 
supersymmetry of the $X\uu + \to - \infty$ theory.\footnote{The analogy is not
precise, however.  The analog of $m\ll{3/2}$ is $\m \cc \exp{\b X\uu +}$,
but there are no physical degrees of freedom that
become heavy in the corresponding limit $X^+ \to \infty$.}   
It is important to note that the worldsheet
supersymmetry is still {\it formally} present, even at infinitely large 
$X\uu +$.  In this limit, however, the supersymmetry is vacuous, 
insofar as its role is simply to eliminate the conformal goldstino degrees of freedom.
The system therefore becomes precisely equivalent to a bosonic string theory.

As noted above (and to be demonstrated explicitly below),
the $X\uu + \to \infty$ limit of our solution is a
dynamical realization of the Berkovits-Vafa embedding of
the bosonic string into the space of solutions of
superstring theory \cite{bv}.  We will thus describe
this limit of our model
in exactly the language and notation of \cite{bv}.
In particular, we will connect our model to that of \cite{bv}
by performing a series of canonical transformations on
the worldsheet fields.  In the next section we describe in detail 
the first stage of these manipulations.

\section{From ultraviolet to infrared variables}
\label{canonical}
We define the indices $\m,\n,\cdots$ to run only over
the light-cone coordinates $\{+,-\}$, and denote the
$D-2$ directions transverse to the light-cone directions
by $i,j,\ldots\,$
Using the canonical flat metric $G\ll{+ -} = G\ll{- +} = -1$,
our Lagrangian for the light-cone multiplets
$X\uu \m,~ \psi\uu\m,~\pst\uu\m$ is:
\bbb
\cllc &= &
{i\over{\pi}}   \pst\uu + \pp\ll + \pst\uu -
+ { i\over{\pi}}   \psi\uu + \pp\ll - \psi\uu -
+ {{i M}\over{2\pi}}  \pst\uu + \psi\uu +
\nn\\ && \nn\\
&&
 - \frac{1}{\pi\alpha'} (\pp\ll + X\uu +)(\pp\ll - X\uu -)
- \frac{1}{\pi\alpha'} (\pp\ll + X\uu - )(\pp\ll - X\uu +)\ ,
\een{newaction}
where $M$ is defined as 
$M \equiv \m\cc\exp{\b X\uu +}$.  For now we will ignore the
transverse degrees of freedom $X\uu i,~\psi\uu i,~ \pst\uu i$.  
Their contributions $G\uu\perp$ and $T\uu\perp$
to the supercurrent and stress tensor are decoupled from the
contributions $\glc$ and $\tlc$
coming from the light-cone degrees of freedom.  In Section \ref{unify}, we will 
re-introduce the transverse degrees
of freedom, but they will play no role until then.
  
The stress tensor of the light-cone sector of the theory is
\bbb
\tlc &=& T\uu{X\uu\m} + T\uu{\psi\uu\m} \ ,
\nn\\\nn\\
T\uu{X\uu\m} &\equiv& - {1\over\apr} G\ll{\m\n} :\pp\ll + X\uu\m \pp\ll + 
X\uu\n : + V\ll\m \pp\ll + \sqd X\uu\m \ ,
\nn\\\nn\\
T\uu{\psi\uu\m} &=&  + {i\over 2}   G\ll{\m\n} :\psi\uu\m \pp\ll + \psi\uu\n :\ ,
\een{stress0}
with supercurrent
\bbb
\glc (\s\uu +) &\equiv& \sqrt{2\over \apr} \psi\ll\m (\pp\ll + X\uu \m)
- \sqrt{2\apr} V\ll\m \pp\ll + \psi\uu\m
\nn\\ && \nn\\
&=& - \sqrt{2\over \apr} \psi\uu + \pp\ll + X\uu -
- \sqrt{2\over \apr} \psi\uu - \pp\ll + X\uu +
+ \sqrt{\apr} ~q~ \pp\ll + \psi\uu + 
+ \sqrt{\apr} ~q~ \pp\ll + \psi\uu - \ .
\een{supercurrent0}
Analogous equations apply for the left-moving stress tensor and
supercurrent, replacing $\psi$ with $\pst$ and $\pp\ll +$
with $\pp\ll -$.

The tensors $\glc(\s)$ and $\tlc(\s)$ are
purely right-moving.  Acting on them with $\pp\ll {\s\uu -}$ yields
a vanishing result, using the definition of the dilaton gradient \rr{vdef}
and the equations of motion
\bbb
 \pp\ll - \psi\uu - &=& \hh M \pst\uu +\ ,
\nn\\ && \nn\\
\pp\ll + \pst\uu - &=& - \hh M \psi\uu +\ ,
\nn\\ && \nn\\
 \pp\ll + \pp\ll - X\uu - &=& - i {{\b \apr}\over 4} M \pst\uu + \psi\uu + \  .
\eee
Written in this manifestly
Lorentz-covariant form, 
\rr{stress0} and \rr{supercurrent0} 
depend on the couplings $\m,\b$
only implicitly.  We can make
the dependence explicit by writing the supercurrent and
stress tensor in terms of canonical variables.
For instance, the fourth
term $\sqrt{\apr}\cc q \cc \pp\ll + \psi\uu -$ in the supercurrent would become
\bbb
\sqrt{\apr}\cc q \cc\lrdd \hh M\tilde{\psi}\uu +
+ \pp\ll{\s\uu 1} \psi\uu - \rrdd\ ,
\nn
\eee
when expressed in canonical variables.



As $M\to \infty$, the massive interaction becomes large
and the theory is strongly coupled in the original variables
$X\uu\m,~ \psi\uu\m,~ \pst\uu\m$.
We would like to define an effective field theory
useful for analyzing the large-$M$ regime,
described by free effective fields whose interactions are
proportional to negative rather than positive powers of $M$.
We cannot, however, derive such a theory in
a conventional, Wilsonian way.
Despite being massive, the coupling $M\pst\uu + \psi\uu +$
does not give rise to massive dispersion relations for the
fermions in the usual sense.
With $M\neq 0$ held constant, the fermions still satisfy
Laplace's equation without mass term.   
The appropriate way to analyze the $M\to \infty$ limit, therefore, 
is not to integrate out degrees of freedom, as
in the examples studied in \cite{previous2}. 
Instead, we will perform a canonical change of variables
such that the new set of variables has interaction
terms {\it inversely } proportional to $M$.
Nothing is integrated out and no information is lost as
$M\to\infty$, but the theory
becomes free in this limit, when expressed in terms of the
new variables.
In what follows, we will motivate and perform the desired change of
variables.


\subsection{Treating $M\cc\pst\uu +\psi\uu +$ as a relevant perturbation}
Let us first consider an approximation in which
the perturbation $M$ is treated as a fixed
constant $M\ll 0$, and only the fermions are treated
as dynamical.
As $M\ll 0 \to\infty$, the conformal invariance of the 
original $\psi\uu\pm,~\pst\uu\pm$ theory is
badly broken.  We would therefore like to find a new
set of variables in which the theory is
approximately conformal, with corrections that
 vanish  in the $M\ll 0\to \infty$ limit.
An appropriate set of variables, which we will dub 
$b\ll 5,~c\ll 5,~\tilde{b}\ll 5,~\tilde{c}\ll 5$,
may be introduced according to
\bbb
\psi \uu + =  2 c\pr\ll 5 - M\ll 0\uu{-1} \tilde{b}\ll 5\ ,
 &\qquad& 
\psi\uu - = M\ll 0 \tilde{c} \ll 5 \ ,
\mmm
\pst\uu + =  - 2  \tilde{c}\pr \ll 5 + M\ll 0\uu {-1} b \ll 5 \ ,
&\qquad& 
\pst\uu - = - M\ll 0 c\ll 5\ ,
\een{zerothcanonical}
where a prime denotes differentiation with respect to $\s\uu 1$.
We are using the equations of motion and the identity $\pp\ll\pm =
\frac{1}{2}(- \pp\ll{\s^0} \pm \pp\ll{\s\uu 1})$.
The change of variables \rr{zerothcanonical} is canonical, but not manifestly
Lorentz invariant.  Applying
\rr{zerothcanonical} to the fermion Lagrangian \rr{newaction}, we obtain
\bbb
{\cal L}\ll{\rm fermi} &= &
- {i\over{\pi}}   \tilde{b} \ll 5 \pp\ll + \tilde{c} \ll 5
- { i\over{\pi}}   b\ll 5 \pp\ll - c \ll 5
- {{i }\over{2\pi M\ll 0 }}  b\ll 5 \tilde{b}\ll 5
\nn\\ && \nn\\
&&
 - \frac{1}{\pi\alpha'} (\pp\ll + X\uu +)(\pp\ll - X\uu -)
- \frac{1}{\pi\alpha'} (\pp\ll + X\uu - )(\pp\ll - X\uu +)\ ,
\een{lag5}
up to a total derivative.  The fermion equations of motion are
\bbb
\pp\ll - c\ll 5 &=& - {1\over{2 M\ll 0}} \tilde{b}\ll 5\ ,
\qquad
\pp\ll + \tilde{c}\ll 5 = + {1\over{2 M\ll 0}} b\ll 5\ ,
\mmm
\pp\ll + \tilde{b}\ll 5 &=& \pp\ll - b\ll 5 = 0 \ ,
\eee
in terms of which the change of variables in Eqn.~\rr{zerothcanonical} is
\bbb
\psi \uu + =  2 \pp\ll + c\ll 5\ ,
 &\qquad &
\psi\uu - = M\ll 0 \tilde{c} \ll 5\ ,
\mmm
\pst\uu + =  2 \pp\ll - \tilde{c}\ll 5\ ,
&\qquad & 
\pst\uu - = - M\ll 0 c\ll 5\ .
\eee

It is now clear that the canonical transformation is
Lorentz invariant if we assign to $b\ll 5$ a Lorentz weight of $3/2$, and
to $c\ll 5 $ a weight of $-1/2$ (with $\tilde{b}\ll 5 $ and $\tilde{c}\ll 5 $ assigned
the opposite Lorentz weights).  Furthermore, if we treat
the nondynamical parameter $M\ll 0$ as having
mass dimension $1$, the fermions can be
assigned the following conformal weights:
\begin{equation}
\begin{array}{|c|c|}
\hline
{\rm field}~&~(\tilde{h}, h) \\
\hline
\hline
b\ll 5  & (0,{3\over 2}) \\
\tilde{b}\ll 5 & ({3\over 2} , 0 ) \\
c\ll 5 & (0 , - {1\over 2} ) \\
\tilde{c}\ll 5 & ( - {1\over 2} , 0 ) \\
\hline
\end{array}
\nn
\end{equation}

We note that our choice of notation $b\ll 5 ,
~c\ll 5 ,~\tilde{b}\ll 5 ,~\tilde{c}\ll 5$ is
not meant to suggest any relationship between our fields and the 
usual reparametrization ghosts.   The present fields, despite being
\zombiesns, have the conventional spin-statistics relation, unlike the
reparametrization
ghosts $b,~c$.  We use the subscript ``$5$'' because we will
eventually carry out a series of
field transformations, labeling each successive set of
canonical variables with a decreasing integer subscript:
in its final incarnation,
the theory will be expressed in terms of the fields
labeled with a subscript ``1.''

We have now shown that the $M\ll 0\to\infty$ limit of a nonunitary pair of Majorana
fermions with a nilpotent mass matrix has a renormalization group flow
to a ghost system with spins $(3/2, -1/2)$.
This is not
the conventional kind of RG flow, insofar as nothing is integrated out.
This flow does have some
conventional properties, however.   For instance, the RG flow
induced by the massive perturbation $M\ll 0 \psi\uu + \pst\uu +$ actually
decreases the central charge by 12 units:  the 
central charge of the original $\psi\uu\pm$ system is $1$, while the central charge
of a $bc$ ghost system with weights $(3/2, -1/2)$ is $-11$.  So, despite the lack of
unitarity, the usual consequence of the $c$-theorem still holds.

\subsection{Promoting $M$ to a dynamical object: $\m\cc\exp{\b X\uu +}$}
Now we want to find a canonical change of variables 
that generalizes \rr{zerothcanonical} to the case for which
$M$ is defined as $\m\cc\exp{\b X\uu +}$, where $X\uu +$ is a
dynamical field.   If we simply carry out the same
change of variables as \rr{zerothcanonical}, we find that the
transformed action has terms bilinear in \zombiesns, multiplying 
derivatives of $X\uu +$.   To eliminate these terms, we need to find corrections
proportional to derivatives of $X\uu +$ to add to the definition of
the new \zombie variables.  We therefore define a {\it new} set of variables 
 $b\ll 4,~c\ll 4 ,~\tilde{b}\ll 4,~\tilde{c}\ll 4$:
\bbb
\psi \uu + &=&  2 c^\prime \ll 4 - M\uu{-1} \tilde{b}\ll 4
+ 2\b (\pp\ll + X\uu +) c\ll 4 \ ,
\nn \\ && \nn \\
\psi\uu - &=& M \tilde{c} \ll 4\ ,
\nn \\ && \nn \\
\pst\uu + &=&  - 2  \tilde{c}^\prime \ll 4 + M\uu {-1} b \ll 4 
+ 2 \b (\pp\ll - X\uu +) \tilde{c}\ll 4 \ ,
\nn \\ && \nn \\
\pst\uu - &=& - M c\ll 4\ .
\een{firstcanonical}

Transforming 
the Lagrangian
classically,
we obtain:
\bbb
\cllc &= &
	- {i\over{\pi}}   \tilde{b} \ll 4 \pp\ll + \tilde{c} \ll 4
	- { i\over{\pi}}   b\ll 4 \pp\ll - c \ll 4
	- {{i }\over{2\pi M }}  b\ll 4 \tilde{b}\ll 4
\nn\\ && \nn\\
&&
	 - \frac{1}{\pi\alpha'} (\pp\ll + X\uu +)(\pp\ll - X\uu -)
	- \frac{1}{\pi\alpha'} (\pp\ll + X\uu - )(\pp\ll - X\uu +)
\nn\\ && \nn\\
&&
	+ {{2 i \b }\over\pi} \m\cc\exp{\b X\uu +}\cc \tilde{c}
	\ll 4 c\ll 4 \cc(\pp\ll + \pp\ll - X\uu +)\ ,
\een{lag4}
up to a total derivative.  To eliminate the
last term, proportional to $\tilde c \ll 4 c\ll 4$, 
we perform a corresponding redefinition of the
bosons $X\uu\pm$:
\bbb
X\uu + &\equiv& Y\uu +\ ,
\mmm
X\uu - &\equiv& Y \uu - + i \b\apr \m\cc\exp{\b X\uu +} c
\ll 4 \tilde{c}\ll 4 \ .
\een{xshift}
This yields the following form of the Lagrangian
\bbb
{\cal L} &= &
- {i\over{\pi}}   \tilde{b} \ll 4 \pp\ll + \tilde{c} \ll 4
- { i\over{\pi}}   b\ll 4 \pp\ll - c \ll 4
- {{i }\over{2\pi M }}  b\ll 4 \tilde{b}\ll 4
\nn\\ && \nn\\
&&
 - \frac{1}{\pi\alpha'} (\pp\ll + Y\uu +)(\pp\ll - Y\uu -)
- \frac{1}{\pi\alpha'} (\pp\ll + Y\uu - )(\pp\ll - Y\uu +)\ ,
\een{lagrangian4}
again up to a total derivative. 

Henceforth we shall refer to the variables
$Y\uu\m,~ b\ll 4,~ c\ll 4,~ \tilde{b}\ll 4,~ \tilde{c}\ll 4$ 
as {\it \ir  variables}, 
and the $X\uu\m,~\psi\uu\m,~ \pst\uu\m$
as {\it \uv variables}.  

The equations of motion for the \ir variables are:
\bbb
\pp\ll + \pp\ll - Y\uu + = 0\ , 
&\qquad&
\pp\ll + \pp\ll - Y\uu - = - {{i \b\apr}\over{4 M}}\cc b\ll 4 \tilde{b}\ll 4 \ ,
\mmm
\pp\ll - c\ll 4 = - {1\over{2 \m}}\cc\exp{- \b Y\uu +}\cc \tilde{b}\ll 4\ ,
&\qquad&
\pp\ll + \tilde{c}\ll 4 = + {1\over{2 \m}}\cc\exp{- \b Y\uu +}\cc b\ll 4\ ,
\mmm
\pp\ll + \tilde{b}\ll 4 = \pp\ll - b\ll 4 = 0 \ . & & 
\eee
Given these, the change of variables can be written in a
manifestly conformally-invariant form:
\bbb
\psi\uu + = 2 \cc(\pp\ll + c\ll 4) + 2\cc\b\cc (\pp\ll + X\uu +) \cc c\ll 4\ ,
&\qquad&
\pst\uu + = 2 \cc(\pp\ll - \tilde{c}\ll 4 ) + 2\cc\b \cc(\pp\ll - X\uu +) \cc\tilde{c}\ll 4\ ,
\mmm
\psi\uu - = \m\cc\exp{\b Y\uu +} \cc\tilde{c}\ll 4 \ ,
&\qquad&
\pst\uu - = - \m \cc\exp{\b Y\uu +}\cc c\ll 4 \ .
\een{lorentzinvar0}
The transformation from \uv variables to
\ir variables is a {\it canonical transformation}:
it takes us from one manifestly canonical set of variables to
another.  This is clear upon inspection of the transformed Lagrangian
\rr{lagrangian4}.  One can also check, working strictly within the Hamiltonian framework,
that the transformation is canonical.  See Appendix A for such a demonstration.

Transforming the stress tensor classically, we obtain:
\bbb
T\uu{Y\uu\m} + T\uu{\psi\uu\m} = 
- \frac{1}{\a'}
G\ll{\m\n}
  \pp_+  Y^\m \pp_+ Y^\n   + V\ll\m \pp\sqd Y\uu\m  
	- \frac{3i}{2}  \pp_+ c\ll 4\, b\ll 4 
	-\frac{i}{2} c\ll 4\, \pp_+ b\ll 4 \ .
\een{classicalstress1}
The \zombie stress tensor is the correct combination to give
weights $-1/2$ and $+ 3/2$ to the $c\ll 4$ and $b\ll 4$
\zombiesns, respectively.  In its manifestly Lorentz-invariant form,
the appearance of the interaction terms is only implicit.  In terms
of the new canonical variables, the interaction terms appear explicitly,
suppressed by $1/M$ (just as they are in the action).  For instance,
in terms of canonical variables, the fermionic stress tensor
is
\bbb
T\uu{\psi\uu\m} =  -  {{3i}\over 2} c'\ll 4 b\ll 4
- {{ i}\over 2} c\ll 4 b'\ll 4 
+ {{3 i }\over {4 M}} \cc \tilde{b}\ll 4 b\ll 4\ .
\eee
As $M$ grows large, the stress tensor becomes
free in canonical variables, with all interaction 
terms going to zero as $M\uu{-1}$.

We now have forms for the action and
the stress tensor that are
manifestly finite in the limit $M\to\infty$.
This indicates that the \ir fields
are legitimate, weakly interacting variables,
suitable for describing the $X\uu +\to +\infty$ limit of the 
theory.  Indeed, there is an exact \it duality \rm
between the \uv description
and the \ir description.
Dualities relating $2D$ 
conformal field theories are often strong/weak dualities,
in the sense that when loop corrections 
in one set of variables are large, loop corrections
in the dual variables are small.  
In the case at hand, loop corrections
are trivial on both sides, and 
the duality
inverts the expansion parameter for \it conformal perturbation theory \rm
rather than for the loop expansion.

We would like to treat the two formulations given above as
exactly describing a single theory.
Even at this stage, however, there is an apparent inconsistency.  
To see this, note that we perturbed the system with
an exactly marginal deformation.
One may rely on diagrammatic reasoning to prove that
the perturbation is exactly marginal, as the theory exhibits no loop 
graphs.  Even so, the central charge appears to have changed in passing
from $X\uu + = - \infty$ to $X\uu + = + \infty$: 
the central charges of the $Y\uu \m$ system and the $X\uu\m$
system would appear to match,\footnote{In particular, the dilaton  
gradient $V\ll + = V\ll - = - {q\over{\sqrt{2}}}$ in the
$X\uu\m,~ \psi\uu\m$ theory 
has not changed in passing to 
the $Y\uu\m,~ b\ll 4,~ c\ll 4$
description by our classical change of variables.}
but the central charge of the fermions has dropped from its original value of
$1$ in the $\psi\uu \pm$ description to the central charge of
$-11$ for a $bc$ ghost system with weights $(3/2, -1/2)$.

To account for the missing central charge,
note that the large-$q$ limit
is, in some sense, a semiclassical limit
of the null Liouville theory.
The amount of central charge that
appears to be missing is subleading in $1 / q\sqd$,
relative to the total central charge of the Liouville
SCFT.  It is therefore a quantum effect.  
This is rather subtle, since the theory has no
nontrivial dynamical Feynman diagrams that might
generate any sort of quantum correction.
Nonetheless, even field theories with 
trivial quantum dynamics can have
quantum renormalizations that affect the conformal properties
of composite operators.  We will see that just such an effect 
accounts for the missing central charge in the $Y\uu\m,~ b\ll 4,~  c\ll 4$ description of
the theory.

\section{Renormalization of the dilaton gradient}
\label{opes}
In this section we will
carefully define normal ordering prescriptions for
composite operators in the \uv and \ir theories.
We will see that the natural normal orderings for the 
\ir variables agree only up to finite terms
with the classical transforms of normal orderings 
for composite operators in the \uv variables.
The effect of these finite differences will
be to renormalize the dilaton gradient of
the system by an amount $\Delta V\ll + =  \b, ~\Delta V\ll - = 0$.
In turn, this effect will add exactly $12$
units of central charge to the linear dilaton theory of $Y\uu\m$.
Much of the detailed manipulation of operators is carried
out in Appendix B.  In this section we will draw on the results
therein to describe the quantum renormalization
of the stress tensor in the transformation to \ir variables.

\subsection{Normal ordering modifications due to interaction terms}
\label{normalorderingdefinitions}

\begin{figure}[htb]
\begin{center}
\includegraphics[width=3.2in,height=0.85in,angle=0]{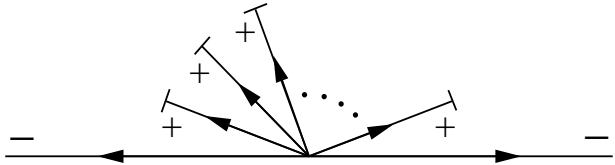}
\caption{Fields on the left-hand side of an OPE correspond
to untruncated external lines in a Feynman diagram,
while fields on the right-hand side correspond to 
truncated external lines.  The ``$+$'' and ``$-$'' labels 
indicate $X^+$ and $X^-$ multiplets, respectively.}
\label{trunc}
\end{center}
\end{figure}

We can learn a great deal about the structure of the
operator product
expansion by looking at the Feynman diagrams of the $2D$ theory.
Every field on the left-hand side of an OPE corresponds
to an untruncated external line in a Feynman diagram,
and every field on the right-hand side of an OPE corresponds
to a truncated external line.  This is depicted in Fig.~\ref{trunc}.
Every connected Feynman diagram in the \uv 
description has exactly one vertex, so any correction to the
free OPE of two fields
must scale with exactly one power of $M$.
Furthermore, the interaction vertices have only ``$-$'' fields as untruncated 
external lines, and only ``$+$'' fields as truncated external lines.  So the
the correction to the
free OPE of two fields
vanishes unless both are ``$-$'' fields.

\ \\

It is easy to see that there must be singularities in
operator products in the interacting theory that are
absent in the free theory.
For example, start with the OPE between $X\uu +(\t)$ and $X\uu -(\s)$:
\bbb
X\uu - (\s) X\uu + (\t) = {{\apr}\over 2} \cc \ln\lba - {1\over{L\sqd}}
(\s\uu + - \t\uu +)(\s\uu - - \t\uu -) \rba
+ :  X\uu - (\s) X\uu + (\t) :\ .
\een{opexplusxminus}
This OPE must be exactly as it is in the 
free theory, since one of the operators in the product is a ``$+$'' field.  
We can use this fact to write a differential equation
for the OPE of $X\uu -(\s)$ with $X\uu -(\t)$.
The differential equation forces a singularity structure for
the $X\uu -(\s) X\uu -(\t)$ 
OPE that differs from that of the free theory.
If we then act on the product with $\pp\ll {\t\uu +} \pp\ll{\t\uu -}$, we obtain
the OPE of $X\uu -(\s)$ with $- i {{\b\apr\m}\over 4}\cc\exp{\b X\uu +(\t)}
\pst\uu + (\t)\psi\uu +(\t)$.  This OPE has a logarithmic singularity as
$\s\to\t$.  We infer that
the OPE of $X\uu -(\s)$ with $X\uu -(\t)$ cannot be completely smooth.

\heading{Normal ordering in the \uv variables}

Using the properties of Feynman diagrams and the equations of motion,
we can derive modified OPEs for the \uv fields.
The natural basis for operators in the \uv
description
is a basis of normal-ordered products
\bbb :X\uu{\m\ll 1}(\r\ll 1)
\cdots X\uu{\m\ll m}(\r\ll m) \cc \psi\uu{\n\ll 1} (\s\ll 1) \cdots\psi\uu{\n\ll n}
(\s\ll n) \pst\uu{\pi\ll 1}(\t\ll 1)
 \cdots 
\pst\uu{\pi\ll p}
(\t\ll p) :\ ,
\een{multilocalop}
with the following properties:
\bi
\item{The normal-ordered operator \rr{multilocalop} is nonsingular when any of the
arguments in the normal ordering symbol approach one another;}
\item{The normal-ordered operators \rr{multilocalop} 
obey the equations of motion.  For instance:
\bbb
\pp\ll {\t\uu +} \pp\ll{\t\uu -} :\cc
X\uu - (\s) X\uu -(\t) \cc:
\cc = - {{i \b \apr \m}\over 4} :\cc X\uu - (\s) \cc\exp{\b X\uu + (\t)} \cc 
\pst\uu + (\t) \psi\uu + (\t) 
:\ ;
\eee }
\item{The normal-ordered product of two ``$+$'' operators is equal to the
ordinary product;}
\item{The normal-ordered product of a ``$+$'' field and a ``$-$'' field is
defined with the subtraction prescription of the free theory;}
\item{The normal-ordered product of two ``$-$'' fields has only ``$+$'' fields
on the right-hand side, and scales as a single power of $M$;}
\item{In the limit $M\to 0$, the structure of the algebra of the
operators \rr{multilocalop} becomes that of the free theory (this property
is implied by the three previous properties).}
\ei
Given these properties, we can derive the full structure of the OPE for
\uv fields in many cases.  We refer the reader to Appendix B for
a more detailed description.

\heading{Normal ordering in the \ir variables}

The normal ordering prescription defined above is not particularly
useful for the \ir
description of the CFT.  The \uv normal ordering $:\cc\cc:$
subtracts
terms from the time-ordered product that are proportional to $M$, which
is very large in the limit where the \ir
variables are free.  

Instead, let us define a second normal ordering prescription, appropriate to the \ir description
of the theory.  In this case we take our basis of operators to be
\bbb \nno Y\uu{\m\ll 1}(\r\ll 1)
\cdots Y\uu{\m\ll m}(\r\ll m) \cc b\ll 4 (\s\ll 1) \cdots b\ll 4
(\s\ll n) \tilde{b}\ll 4 (\t\ll 1)
 \cdots 
\tilde{b}\ll 4
(\t\ll p) c\ll 4 (\zeta\ll 1)\cdots c\ll 4 (\zeta \ll q) \tilde{c}\ll 4 (\o\ll 1)\cdots
\tilde{c}\ll 4 (\o\ll r) \nno, \
\een{multilocalop2}
which have the following properties:
\bi
\item{The normal-ordered operator \rr{multilocalop2} is nonsingular when any of the
arguments of operators in the normal ordering symbol approach one another;}
\item{The normal-ordered operators \rr{multilocalop2} obey the equations of motion.  
For instance:
\bbb
\pp\ll {\t\uu +} \pp\ll{\t\uu -} \nno \cc
Y\uu - (\s) Y\uu -(\t) \cc \nno 
\cc = - {{i \b \apr }\over {4\m}} \nno \cc Y\uu - (\s) \cc\exp{- \b Y\uu + (\t)} \cc 
b\ll 4  (\t) \tilde{b}\ll 4 (\t)
\nno\ ;
\eee }
\item{The normal-ordered product of two  operators from
the set $b\ll 4 ,\tilde{b}\ll 4 , Y\uu +$ is equal to the
ordinary product;}
\item{The normal-ordered product of a 
field from the set
$c\ll 4, \tilde{c}\ll 4, Y\uu -$ with a field from the
set $b\ll 4, \tilde{b}\ll 4, Y\uu +$ is 
defined with the subtraction prescription of the free theory;}
\item{The normal-ordered product of two fields
from the set
$c\ll 4, \tilde{c}\ll 4, Y\uu -$ 
 has only fields from the set $b\ll 4, \tilde{b}\ll 4, Y\uu +$ 
on the right-hand side, and scales as a single power of $M\uu{-1}$;}
\item{In the limit $M\to \infty$, the structure of the algebra of the
operators \rr{multilocalop2} becomes that of the free theory of 
the \ir fields (this property
is implied by the three previous properties).}
\ei

\subsection{Operator product expansions in the interacting CFT}
In Appendix B we solve the structure of several OPEs that are necessary
for defining composite operators, such as
the stress tensor and supercurrent.
Here, we list some of the results.  In some cases we will need only the singular
terms in the OPE, and in some cases we will need some of the smooth terms
in the OPE as well.  For the OPEs involving $X\uu - (\s)$ with $\psi\uu - (\t)$ and 
$\pst\uu - (\t)$, we have:
\bbb
X\uu - (\s) \psi\uu -(\t)  & \simeq &
: X\uu - (\s) \psi\uu -(\t) :
\mmm
&&
\kern-30pt
+{{\b\apr}\over 4} \ln\lsqq - {1\over{L\sqd}}
(\s\uu + - \t\uu + )(\s\uu - - \t\uu - ) \rsqq
\int\ll{\s\uu -}\uu{\t\uu -}
dy M(\t\uu +, y) \pst\uu + (y) \ ,
\een{opexminuspsiminus}
and
\bbb
X\uu - (\s) \pst\uu -(\t) &\simeq&
: X\uu - (\s) \pst\uu -(\t) :
\mmm
&&
\kern-30pt
- {{\b\apr}\over 4} \ln\lsqq - {1\over{L\sqd}}
(\s\uu + - \t\uu + )(\s\uu - - \t\uu - ) \rsqq
\int\ll{\s\uu +}\uu{\t\uu +}
dy M(y, \t\uu -) \psi\uu + (y) \ ,
\een{opexminuspsitildeminus}
where the $\simeq$ denotes equality modulo smooth terms.

The OPEs of $Y\uu -$ with $c\ll 4$ and $\tilde{c}\ll 4$ are given by:
\bbb
Y\uu - (\s) c\ll 4(\t) &\simeq&
\nno Y\uu - (\s) c\ll 4(\t) \nno
\mmm
&& \kern-25pt
+ {{\b\apr}\over 4} \ln\lsqq - {1\over{{L^{\prime 2}}}}
(\s\uu + - \t\uu + )(\s\uu - - \t\uu - ) \rsqq
\int\ll{\s\uu -}\uu{\t\uu -}
dy N(\t\uu +, y) \tilde{b}\ll 4 (y) \ ,
\een{opeyc}
\bbb
 Y\uu - (\s) \tilde{c}\ll 4 (\t) &\simeq&
\nno Y\uu - (\s) \tilde{c}\ll 4 (\t) \nno
\mmm
&&
\kern-25pt
- {{\b\apr}\over 4} \ln\lsqq - {1\over{L\uu{\prime 2}}}
(\s\uu + - \t\uu + )(\s\uu - - \t\uu - ) \rsqq
\int\ll{\s\uu +}\uu{\t\uu +}
dy N(y, \t\uu -) b\ll 4 (y) \ ,
\een{opeyctilde}
where
\be
N(\s) \equiv M\uu{-1} (\s) = {1\over\m} \exp{- \b X\uu +}\ .
\ee

For these operator products we will never need the smooth terms, so we do
not compute them.
In Appendix B we also derive the $\psi\uu - (\s)\pst\uu -(\t)$ and
$c\ll 4 (\s) \tilde{c}\ll 4 (\t)$ OPEs, with smooth terms included.
We find: 
\bbb
\psi \uu - (\s) \pst\uu - (\t) &=& : \psi \uu - (\s) \pst\uu - (\t) :
- { i \over 2} \ln \lba {1\over{L\sqd}} (\t\uu + - \s\uu +) (\t\uu - -
\s\uu -) \rba M(\s\uu +, \t\uu -) 
\mmm
&&
\kern-20pt
- {i\over 2} \sum\ll{n = 1}\uu\infty {1\over{n\cdot n!}} \lsqq
(\t\uu + - \s\uu +)\uu n \cc \pp\ll +\uu n + (\s\uu - - \t\uu -)\uu n \cc
\pp\ll - \uu n \rsqq
M(\s\uu + , \t\uu -)\ .
\een{totalminusminusope}
and
\bbb
 c\ll 4 (\s) \tilde{c}\ll 4 (\t) 
 &=& 
\nno c\ll 4 (\s) \tilde{c}\ll 4 (\t) \nno
- { i \over 2} \ln\lba {1\over{L\psq}} (\t\uu + - \s\uu +) (\t\uu - -
\s\uu -) \rba N(\s\uu +, \t\uu -)
\mmm
&&
\kern-10pt
 - {i\over 2} \sum\ll{n = 1}\uu\infty {1\over{n\cdot n!}} \lsqq
(\t\uu + - \s\uu +)\uu n \cc \pp\ll + \uu n + (\s\uu - - \t\uu -)\uu n \cc
\pp\ll - \uu n \rsqq
N(\s\uu + , \t\uu -)\ .
\eee

\subsection{Quantum correction to the \ir stress tensor}
Given our normal ordering definitions, we can now
apply the change of variables in Eqns.~\rr{xshift} and \rr{lorentzinvar0} to
composite operators.  Starting with
composite expressions in the fields of the $X,~\psi$ theory, we
will derive corresponding expressions in the $b_4,~c_4,~Y$
theory, including quantum corrections.
The bosonic stress tensor turns out to
transform unproblematically,
while the fermionic stress tensor picks up a quantum correction due to
the mismatch between $:\cc\cc:$ and $\nno\cc\cc\nno$ normal ordering
prescriptions.

In Appendix B we derive the quantum-corrected
transformations of 
the various components of the stress tensor.
The result (Eqn.~\rr{quantumstress2}) is
simply equal to the classical result
\rr{classicalstress1}, ordered with $\nno\cc\cc\nno$
normal ordering, and with the addition of a quantum correction:
\bbb
\tlc &\equiv &T\uu{X\uu\m} + T\uu{\psi\uu\m}
\mmm
&&
\kern-00pt
= - \frac{1}{\a'}
G\ll{\m\n}
  \nno \pp_+  Y^\m \pp_+ Y^\n\nno    + \hat{V}\ll\m \pp\ll + \sqd Y\uu\m  
	- \frac{3i}{2}  \nno \pp_+ c\ll 4\, b\ll 4 \nno 
	-\frac{i}{2} \nno c\ll 4\, \pp_+ b\ll 4 \ \nno \ ,
\een{quantumstress3}
where we define $\hat{V}$ to be the renormalized dilaton gradient:
\bbb
&&
\hat{V}\ll\m \equiv
V\ll\m + \Delta V\ll\m\ ,
\mmm
&&
\Delta V\ll + = + \b\ , \qquad  \Delta V\ll - = 0\ .
\een{vhatdef}
Our total stress tensor in \ir variables
turns out to be the classical piece in Eqn.~\rr{classicalstress1}
plus the quantum correction $\Delta V\ll\m \pp_+\sqd Y\uu\m$.
The quantum correction is of order 
$O({{\b}\over q}) = O(\b\sqd) = O(q\uu{-2})$ relative
to the classical term.  


Of course, the renormalization of the dilaton contributes to
the central charge.  Since the metric is
unrenormalized, we are left with a
contribution to the central charge equal to
\bbb
c\uu{\rm dilaton} = 6\apr \eta\uu{\m\n} \hat{V}\ll \m
\hat{V}\ll\n = - 6\apr q\sqd + 6 \sqrt{2} \apr \b q\ .
\eee
Using the values for $\b$ and $q$ found in Eqns.~\rr{vdef} and \rr{betaqrel}, 
we have 
\bbb
c\uu{\rm dilaton} =  12 - \frac{3}{2}(D - 10) = 27 - {{3D}\over 2}\ .
\een{newcdilaton} 
The central charge contribution of the free fields is $2$ from the $Y\uu\m$, 
$-11$ from the $b\ll 4 c\ll 4$ system, and
${3\over 2} (D - 2) $ from the transverse degrees of freedom
$X\uu i,~ \psi\uu i$,
 for a total free-field contribution of ${{3D}\over 2} - 12$.
This demonstrates that
the total central charge in the theory is always equal
to $15$.  As one moves in the target space from $X\uu + = - \infty$
to $X\uu + = + \infty$, twelve units of central charge are
transferred from the lightcone fermions $\psi\uu\pm$ to the
dilaton gradient.  This is the effect of {\it central charge transfer} that 
was discussed in \cite{hellerman2,previous2,freedman},
but operating by a different mechanism.  The central charge being transferred to
the dilaton gradient does not occur through a loop diagram of massive fields 
being integrated out.
Instead, the central charge is transferred through a mismatch of
normal ordering prescriptions appropriate to the free field theories in 
the two limits $X\uu +\to\pm\infty$.

\subsection{Quantum correction to the \ir supercurrent}
Using the same method we used to transform 
the stress tensor, we can calculate the
transformation of the supercurrent
from $\psi,~X$ variables to $b_4,~c_4,~Y$ variables,
including quantum corrections.  Starting from
Eqn.~\rr{supercurrent0}, we find that three of the
four pieces transform classically, and the
fourth acquires a quantum correction.  For convenience of presentation, 
we define the following as terms in the supercurrent 
(i.e.,~$G^{\rm LC} = {\bf 1} + {\bf 2} + {\bf 3} + {\bf 4}$):
\bbb
{\bf 1} \equiv
- \sqrt{2\over \apr} \psi\uu + (\pp\ll + X\uu  -)\ ,
&\qquad&
{\bf 2} \equiv  - \sqrt{2\over \apr} \psi\uu - (\pp\ll + X\uu  +) \ ,
\mmm
{\bf 3} \equiv \sqrt{\apr}\cc q \cc \pp\ll + \psi\uu + \ ,
&\qquad &
{\bf 4} \equiv  \sqrt{\apr}\cc q \cc \pp\ll + \psi\uu - \ .
\een{supercurrentpieces}
We find that the classical transformation of
term ${\bf 1}$ is
\bbb
{\bf 1}\ll{\rm classical} 
= - 2 \sqrt{2\over{\apr}}
\nno \lsqq (\pp\ll + c\ll 4) (\pp\ll + Y\uu -)
+ \b c\ll 4 (\pp\ll + Y\uu +) (\pp\ll + Y\uu -) 
- {{i \b\apr}\over 2}  (\pp\ll + c\ll 4) b\ll 4 c\ll 4 
 \rsqq \nno\ ,
\eee
while the quantum correction takes the form
\bbb
{\bf 1}\ll{\rm quantum}
= - \b\sqrt{{{\apr}\over 2}}
 \pp\ll +^2 c\ll{4}
-2 \b\sqd \sqrt{ {{ \apr}\over 2}}
c\ll 4 \pp\ll +^2 Y\uu + \ .
\eee
In the classical transformation we have made use of
fermi statistics and the equation of motion for $\tilde{c}_4$.

The other three terms are given by classical
substitution,
because the corresponding operator products have no
singularities coming from the interaction terms:
\bbb
{\bf 2} &=& - \sqrt{2\over\apr}
M (\pp\ll + Y\uu +) \tilde{c}_4\ ,
\mmm
{\bf 3} &=& 
2 q \sqrt{\apr} (\pp\ll +\sqd c\ll 4) + 2 \b q \sqrt{\apr} 
(\pp\ll + Y\uu +)(\pp\ll + c\ll 4) + 2 \b q \sqrt{\apr} c\ll 4 (\pp\ll +
\sqd Y\uu +)\ ,
\mmm
{\bf 4}  &=& q \sqrt{\apr} M~(\pp\ll + \tilde{c}\ll 4)
+ q\sqrt{\apr}  \b \tilde{c}_4~ M (\pp\ll + Y\uu +)\ .
\eee
Using the equations of motion for $\tilde{c}_4$ and
the relation $\b q = {{\sqrt{2}}/{\apr}}$, we are
left with
\bbb
{\bf 1}\ll{\rm classical}
&=& - 2 \sqrt{2\over{\apr}}
\nno \lsqq (\pp\ll + c\ll 4) (\pp\ll + Y\uu -)
+ \b c\ll 4 (\pp\ll + Y\uu +) (\pp\ll + Y\uu -) 
- {{i\b\apr}\over 2}  (\pp\ll + c\ll 4) b\ll 4  c\ll 4   \rsqq \nno\ ,
\mmm
{\bf 1}\ll{\rm quantum} &=& - \b\sqrt{{{\apr}\over 2}}
 \pp\ll +^2 c\ll{4}
-2 \b\sqd \sqrt{ {{ \apr}\over 2}}
c\ll 4 \pp\ll +^2 Y\uu + \ ,
\mmm
{\bf 2} + {\bf 4} &=&  {q\over 2}  \sqrt{\apr} b\ll 4\ ,
\mmm
{\bf 3} &=& 
2 q \sqrt{\apr} (\pp\ll +\sqd c\ll 4) + 2 \sqrt{2\over\apr}
(\pp\ll + Y\uu +)(\pp\ll + c\ll 4) + 2 \sqrt{2\over\apr} c\ll 4 (\pp\ll +
\sqd Y\uu +)\ .
\een{finalsupercurrent1}


It is straightforward to verify that
the supercurrent $\glc\equiv {\bf 1}
\ll{\rm classical} + {\bf 1}
\ll{\rm quantum} + {\bf 2} + {\bf 3} + {\bf 4}$,
as written in the $b\ll 4,~c\ll 4,~Y$
variables in Eqn.~\rr{finalsupercurrent1},
indeed closes on the stress tensor in Eqn.~\rr{quantumstress3}:
\bbb
\glc(\s) \glc(\t)  &=&
 {{2 i }\over {3(\t\uu + - \s\uu +)\uu 3}}~c\uu{\rm LC}
+ {{2 i }\over{\t\uu + - \s\uu +}} \tlc(\t) 
+ \rm (finite\ terms)\ ,
\een{virasoroope}
where
\bbb
\clc\equiv 3 + c\uu{\rm dilaton}\ ,
\eee
with
$ c\uu{\rm dilaton}$ given in \rr{newcdilaton}.

One important feature of the supercurrent is
that, expressed in $b\ll 4,~c\ll 4,~Y $ variables, 
it is manifestly finite in the limit $X\uu + \to + \infty$
(as is the stress tensor).  This completes the proof
that the $b\ll 4,~c\ll 4,~Y $ fields can be regarded as dual variables
that render the theory free in the $M\to\infty$ limit.

\section{Equivalence to the Berkovits-Vafa embedding}
\label{unify}
We have found an initial canonical transformation
that takes us from the \uv
degrees of freedom, in terms of which the $X\uu + \to - \infty$
limit of the theory is free, to the \ir system, 
which is free in the $X\uu + \to +\infty$ limit.
We now wish to understand the
physics of the $X\uu +  \to + \infty$ limit on its
own terms.  We will focus strictly on the 
limiting regime of the \ir theory.   In practice, this
means that, when written in \ir variables,  
we discard the $\exp{- \b Y\uu + }\tilde{b}\ll 4
b\ll 4$ term in the action, as
well as any $\exp{- \b Y\uu +}$ terms
in the supercurrent and stress tensor. 
Furthermore, in the $M\to \infty $ limit, $\pp_+ c_4 = c_4'$, and we will
use both notations interchangeably.  (However, we will usually use
$\pp\ll + c_4 $ when it clarifies the Lorentz properties of an expression.)
Since the field theory is free in these variables,
the right- and left-moving fields decouple.  We are therefore
able to consider the right-moving fields
$b\ll 4,~c\ll 4$ alone, since the discussion of their
dynamics applies trivially to the
$\tilde{b}\ll 4,~\tilde{c}\ll 4$ fields.

\subsection{Second canonical transformation and manifest $\mathbb{Z}\ll 2$ reflection symmetry}
For later convenience, we will rescale the $b\ll 4$ \zombie so that 
the new $b$ fermion appears in the supercurrent with unit normalization.
To preserve all canonical commutators, however, we will rescale
the $c_4$ field oppositely:
\bbb
b\ll 4 =  \frac{2}{q\sqrt{\apr}}b_3  =  \b \sqrt{2\apr}\, b_3 \ ,
\xxx
c\ll 4 =  \frac{q\sqrt{\apr}}{2} c_3 = \frac{1}{\b \sqrt{2\apr}} c_3  \ .
\een{inverttform}
In terms of the $c\ll 3,~b\ll 3$ fields, the
supercurrent appears as:
\bbb
 {\bf 1}_{\rm classical} 
		& = & - \frac{2}{\apr} \nno \left( \frac{1}{\b } \pp_+ c_3 
		\pp_+ Y^-  + c_3 \pp_+ Y^+ \pp_+ Y^- 
		 -\frac{i\apr}{2} \pp_+ c_3 b_3 c_3 \right)\nno \ ,
\mmm
{\bf 2} + {\bf 4} & = & b_3   \ ,
\mmm
{\bf 3} + {\bf 1}_{\rm quantum} & = & \left( q^2 \apr - \frac{1}{2}\right) \pp_+^2 c_3
	+ \frac{2}{\b \apr} \pp_+ c_3 \pp_+ Y^+ 
	+ \left( \frac{2}{\b\apr} - \b \right) c_3 \pp_+^2 Y^+ \ .
\eee
It is straightforward to see that the 
stress tensor and action are unchanged:
\bbb
\cllc &= &
- {i\over{\pi}}   \tilde{b} \ll 3 \pp\ll + \tilde{c} \ll 3
- { i\over{\pi}}   b\ll 3 \pp\ll - c \ll 3
\mmm
&&
 - \frac{1}{\pi\alpha'} (\pp\ll + Y\uu +)(\pp\ll - Y\uu -)
- \frac{1}{\pi\alpha'} (\pp\ll + Y\uu - )(\pp\ll - Y\uu +)\ ,
\mmm
\tlc & = & 
- \frac{1}{\a'}
G\ll{\m\n}
  \nno \pp_+  Y^\m \pp_+ Y^\n\nno    + \hat{V}\ll\m \pp\ll + \sqd Y\uu\m  
	- \frac{3i}{2}  \nno \pp_+ c\ll 3\, b\ll 3 \nno 
	-\frac{i}{2} \nno c\ll 3\, \pp_+ b\ll 3 \ \nno \ .
\eee

The invariance properties of the system
under spatial reflection are still
unclear.  The stress tensor
is invariant under the discrete symmetry reflecting
the spacelike vector orthogonal to
$\hat{V}\ll\m$.  The supercurrent
is not, however, since $V\ll\m$ and $\Delta V\ll \m$
appear independently in $\glc$.
We would like to find field variables that render
this discrete symmetry more manifest, such that only the vector
$\hat{V}\ll\m$ enters $\glc$.

We therefore define new variables $b\ll 2,~c\ll 2,~Z\uu\m$ by:
\bbb
Y\uu \pm &= & Z\uu\pm \pm {i\over{2\b}} c_2 \partial_+ c_2\ ,
\mmm
b_3 &=&  b\ll 2 - {2\over{\b\apr}} (\pp\ll + c_2)\lrdd
    \pp\ll + Z\uu + - \pp\ll + Z\uu - \rrdd
 - {1\over{\b\apr}} c_2 \lrdd \pp\sqd\ll + Z\uu +
- \pp\ll + \sqd Z\uu - \rrdd
\mmm
&& + {i\over{2\b\sqd \apr}} c_2(\pp\ll + c_2)(\pp\sqd\ll + c_2)\ ,
\mmm
c_3 &=& c_2\ .
\een{threestotwos}
As previously noted, one may verify that this change of variables
is canonical either by checking commutators directly,
or by noting that the variable change can be 
derived by exponentiating an infinitesimal transformation.
In terms of the infinitesimal generator
\bbb
h\equiv \frac{i}{2 \pi \b  \apr} \int d\s_1  
	\left( \pp_+ Y^+ - \pp_+ Y^- \right) c_3\, \pp_+ c_3\ ,
\eee
we define our new variables via
\bbb
Z\uu\m  &\equiv&  \exp{{i} h} \cc Y^\m \cc \exp{-i h} \ ,
\mmm
b\ll 2  &\equiv&   \exp{i h} \cc b_3 \cc \exp{-i h}\ , 
\mmm
c\ll 2  &\equiv&   \exp{i h} \cc c_3 \cc \exp{-i h}\ .
\een{transform2}
Carrying out the transformation explicitly, we recover the replacement rules
\bbb
Z\uu\pm &=& Y^\pm \mp \frac{i}{2\b} c_3 \pp_+ c_3 \ ,
\mmm
b\ll 2  & = & b_3 + \frac{2}{\b\apr} \left( \pp_+ Y^+ 
	- \pp_+ Y^- \right) \pp_+ c_3 + \frac{1}{\b \apr} \left( \pp_+^2 Y^+ 
	- \pp_+^2 Y^- \right) c_3
\mmm
&&	+ \frac{i}{2 \b^2 \apr}c_3 \pp_+ c_3 \pp_+^2 c_3\ ,
\mmm
c\ll 2 & = & c_3\ ,
\eee
which invert the expressions in Eqn.~\rr{threestotwos}.
It follows that the transformation in Eqn.~(\ref{threestotwos}) is
indeed canonical.

In terms of $b\ll 2,~c\ll 2$ and $Z\uu\m$ variables, the 
action on a flat worldsheet is completely unchanged (modulo
the replacement $b_3 \to b_2,~c_3 \to c_2$ and $Y^\m \to Z^\m$).
Furthermore, the supercurrent now preserves the $\mathbb{Z}\ll 2$ little group
of the linear dilaton:
\bbb
\glc(\s) &=&  b_2 + i \pp_+ c_2 b_2 c_2 - c_2 T^{Z^\m} 
	+ \left( \apr q^2 - \frac{1}{2}\right)
	\pp_+^2 c_2 
\mmm
&&	- \left( \frac{i}{4} \apr q^2 + \frac{i}{2} \right)
	c_2 \pp_+ c_2 \pp_+^2 c_2\ .
\een{supercurrent2}
The stress tensor acquires an additional term
bilinear in the $c\ll 2$ field:
\bbb
\tlc &=&  T^{Z^\m } + T^{b_2 c_2} \ ,
\mmm
T^{Z^\m} & \equiv & -\frac{1}{\apr} \nno G_{\m\n}\pp_+Z^\m \pp_+ Z^\n \nno 
	+ \hat V_\m \pp_+^2 \hat V^\m \ ,
\mmm
T^{b_2 c_2} & \equiv & -\frac{3i}{2}  \nno \pp_+ c\ll 2\, b\ll 2 \nno 
	-\frac{i}{2} \nno c\ll 2\, \pp_+ b\ll 2 \ \nno 
	+ \frac{i}{2}\pp_+^2 \left( c_2 \pp_+ c_2 \right)\ .
\een{stresstensor2}

Since the transformation in Eqn.~\rr{transform2} preserves all commutators,
it follows that the OPE in Eqn.~\rr{virasoroope} still holds.
Our superconformal theory now appears as a sum
of two disjoint sectors.  In the sector involving 
$Z\uu\pm,~b\ll 2,~c\ll 2$, the supercurrent is $\glc $ and
the stress tensor is $\tlc$.
In the other sector, the supercurrent and stress tensor take the form
\bbb
G^\perp  & \equiv & \sqrt{ \frac{2}{\apr}} \psi^i \pp_+ X_i\ ,
\qquad i \in 2, \ldots, D-1    \ ,
\mmm
T^\perp & \equiv & -\frac{1}{\apr} : \pp_+ X\ll i \pp_+ X \ll i : 
	\, + \, \frac{i}{2} : \psi^i \pp_+ \psi^i  :\ .
\eee
Worldsheet supersymmetry is now realized nonlinearly, in the sense that
there is a goldstone fermion $c\ll 2 $,
and fields are no longer organized into multiplets.  
Instead, the bosons $Z\uu \m$ transform into their own
derivatives, times a goldstone field:
\bbb
[Q,Z\uu\m ]  =  i c_2 \pp_+ Z^\m   \ ,
\qquad  ~~
\{Q, c\ll 2 \}  =  1 + i c_2 \pp_+ c_2   \ ,
\eee
where
\bbb
Q\equiv \frac{1}{2\pi} \int d\s\ll 1 \cc G(\s)\ .
\eee
This is exactly the universal nonlinear realization
of supersymmetry first described by Volkov and Akulov
\cite{va}.  In the sector involving the transverse
fields $X\ll i,~\psi\uu i$, supersymmetry is
realized in the usual linear fashion:
\bbb
[Q, X\ll i] &=& i\sqrt{ \frac{\apr}{2}} \psi^i \ ,
\mmm
\{Q, \psi\uu i \} &=& \sqrt{\frac{2}{\apr}}\pp_+ X\ll i \ .
\eee
At first sight, our realization of supersymmetry
in the full theory is
unfamiliar, with worldsheet
supersymmetry realized linearly in one sector and
nonlinearly in another.  However, we shall now see that 
in a conformal field theory, such a realization is
equivalent to one in which worldsheet supersymmetry is
realized completely nonlinearly in \it all \rm sectors.

\subsection{Final canonical transformation and manifest $SO(D-1)$}
Until now, we have largely ignored $T\uu{\perp}$ and $G\uu{\perp}$.  
They are the usual free stress tensor and
supercurrent for $D - 2$ free massless multiplets $X\ll i,~\psi\uu i$.  
$T\uu{\perp}$ and $G\uu{\perp}$ obey the standard
equal-time commutation
relations for a SCFT of central charge $c\uu\perp
\equiv {3\over 2} (D - 2)$:
\bbb
\left[T\uu\perp(\s) , T\uu{\perp}(\t)\right] &=&  
	-2\pi i\, \delta (\s_1 - \t_1) \pp_+ T^\perp (\t)
	+ 4 \pi i \,\delta'(\s_1 - \t_1) T^\perp(\t)
	+ \frac{ \pi i }{6}c^\perp \,\delta'''(\s_1 - \t_1)    \ ,
\mmm
\left[T\uu\perp (\s) , G\uu \perp (\t) \right] &=&  
	- {2\pi i}\,\delta (\s_1 - \t_1) \pp_+ G^\perp (\t) 
	+ 3\pi i \,\delta' (\s_1 - \t_1 ) G^\perp (\t)      \ ,
\mmm
\left[ G\uu\perp (\s) , T\uu\perp (\t)  \right] &=& 
              	- \pi i \,\delta  (\s_1 - \t_1 ) \pp_+ G^\perp (\t)
	      + 3\pi i \,\delta' (\s_1 - \t_1 ) G^\perp (\t )\ ,
\mmm
\{ G\uu\perp (\s) , G\uu\perp (\t)  \} &=& 
	  -4\pi \,\delta(\s_1 - \t_1) T^\perp (\t)
	-\frac{2\pi}{3} c^\perp \,\delta'' (\s_1 - \t_1)\ .
\een{tgcomms}

We now perform a final transformation on the system.
Defining the Hermitian infinitesimal generator
\bbb
g\equiv  - {i\over{2\pi}} \int d\s\ll 1 c\ll 2 (\s) G\uu\perp (\s)\ ,
\eee
we transform all operators in the theory according to
\bbb
\co \to U\cc \co \cc U\uu{-1}\ ,
\eee
with $U\equiv \exp{i g}$.  We therefore obtain
\bbb
b\ll 1 \equiv U\uu{- 1} \cc b\ll 2 \cc 
U\uu{+ 1}  \ ,
& \qquad &
c\ll 1 \equiv U\uu{- 1} \cc c\ll 2 \cc 
U\uu{+ 1}  = c\ll 1 \ , 
\mmm
Z\uu i \equiv U\uu{- 1} \cc X\uu i \cc
U\uu{+ 1}  \ ,
& \qquad &
\psi\uu{Z\uu i} \equiv U\uu{- 1} \cc
\psi\uu i \cc U\uu{+ 1} \ ,
\mmm
\hat{G}\uu\perp \equiv U\uu{- 1}\cc
G\uu\perp \cc U\uu{+ 1}  \ , 
& \qquad &
\hat{T}\uu\perp \equiv U\uu{- 1}\cc
T\uu\perp \cc U\uu{+ 1}  \ .
\eee

Clearly the $b\ll 1,~c\ll 1,~Z\uu i ,~\psi\uu{Z\uu i}$ variables
are a canonical set.  In particular, $\hat{G}\uu\perp$
and $\hat{T}\uu\perp$ can be seen to obey the same
equal-time commutation relations (cf.~Eqns.~\rr{tgcomms})
as $T$ and $G$.
Since $[g,g] = 0$, we can rewrite the generator $g$ in terms
of the new variables as
\bbb
g &=& U\uu{-1} \cc g \cc U\uu{+1}
\mmm
&=& - {i\over{2\pi}} \int d\s\ll 1 \cc c\ll 1(\s) \hat{G}\uu\perp (\s) \ .
\eee
We can therefore invert the transformation to obtain
\bbb
b\ll 2 &=&  U\uu{+ 1} b\ll 1 U\uu{- 1} =  b_1 - \hat G^\perp
	+ c_1 \hat T^\perp + \frac{1}{6} c^\perp c_1''
	-\frac{i}{2} c_1 c_1' \hat G^\perp
	+ \frac{i}{24} c^\perp c_1 c_1' c_1''\ ,
\mmm
c\ll 2 &=& U\uu{+ 1} c\ll 1 U\uu{- 1}  = c_1 \ ,
\mmm
X\ll i &=& U\uu{+ 1} Z\uu i U\uu{- 1} = Z_i + i c_1 \psi^{Z^i} \ ,
\mmm
\psi\uu i &=& U\uu{+ 1} \psi\uu{Z\uu i}  U\uu{- 1} = \psi^{Z^i} + 
	c_1 \pp_+ Z^i + \frac{i}{2} c_1 c_1' \psi^{Z^i}  \ ,
\mmm
G\uu\perp &=& U\uu{+ 1} \hat{G}\uu\perp U\uu{- 1} = 
	\hat G^\perp -2 c_1 \hat T^\perp
	-\frac{1}{3} c^\perp c_1''
	+\frac{3i}{2} c_1 c_1' \hat G^\perp
	-\frac{i}{6} c^\perp c_1 c_1' c_1''  \ ,
\mmm
T\uu\perp &=& U\uu{+ 1} \hat{T}\uu\perp U\uu{- 1} = \hat T^\perp 
	- \frac{3i}{2} c_1' \hat G^\perp - \frac{i}{2} c_1 \pp_+ \hat G^\perp
	+ \frac{i}{4} c^\perp c_1' c_1''
	+ \frac{i}{12} c^\perp c_1 c_1'''  
- i \, T^\perp c_1 c_1' 
\ . 
\mmm
& &
\eee
Using these, we compute the transformation of the following
quantity, which enters the stress tensor for the $b_2$ and $c_2$ \zombiesns.
\bbb
-\frac{3i}{2} \nno c_2' b_2 \nno 
	- \frac{i}{2} \nno c_2 b_2' \nno     
& = & U\uu{+ 1} \left(-\frac{3i}{2} \nno c_1' b_1 \nno 
	- \frac{i}{2} \nno c_1 b_1' \nno    
	\right) U\uu{- 1} 
\mmm
	&=& -\frac{3i}{2} \nno c_1' b_1 \nno  - \frac{i}{2} \nno c_1 b_1' \nno
	+ \frac{3i}{2} c_1' \hat G^\perp
	- \frac{i}{4}c^\perp c_1' c_1''
\mmm
&&
	- \frac{1}{12}c^\perp c_1 c_1'''
	+ i c_1 c_1' \hat T^\perp
	+ \frac{i}{2} c_1 \pp_+ \hat G^\perp\ .
\eee
The total supercurrent $G \equiv \gzpmbc + G\uu\perp$ is then
\bbb
G &=& b_1 + i \nno c_1' b_1 c_1 \nno - c_1 T^{\rm mat} + c_1'' \left(-\frac{1}{6}c^\perp
	-\frac{1}{2} + \apr q^2 \right) 
\mmm
&&	+ c_1 c_1' c_1'' \left(
	-\frac{i}{4} \apr q^2 - \frac{i}{2} + \frac{i}{24}c^\perp
	\right)  \ .
\eee
We have defined 
\bbb
T\uu{\rm mat} \equiv T^{Z^\pm} + T^{Z^i} + T^{\psi^i} + \hat T^\perp 
\eee
as the standard stress tensor for $D-2$ fermions $\psi\uu i$
and $D$ bosons $Z\uu M$ ($M \in \{\pm,i\}$), with a linear dilaton
\be
\hat{\Phi} \equiv  \hat V \cdot Z \ .
\ee
The total transformed stress tensor is the sum
\bbb
\tfif = T\uu{\rm mat} + T\uu{b_1 c_1}\ ,
\eee
and the total supercurrent and stress tensor still
satisfy the OPE in Eqn.~\rr{virasoroope}, as they must.
The $b_1 c_1$ stress tensor takes the explicit form
\be
T^{b_1 c_1}  &= & 
	-\frac{3i}{2} \nno \pp_+ c_1 b_1 \nno
	-\frac{i}{2} \nno c_1 \pp_+ b_1 \nno
	+ \frac{i}{2} \pp_+ (c_1 \pp_+^2 c_1)\ .
\label{tbcstress}
\ee

For the actual values of the parameters in our solution,
$q = \sqrt{{D - 10}\over{4\apr}}$ and $c\uu{\perp} = {3\over 2} (D - 2)$, 
the term trilinear in $c_1$ drops out of the supercurrent, and the 
coefficient of $c\ll 1\prpr$ is always $-{5\over 2}$,
independent of $D$:
\bbb
G = b_1 + i \nno c_1' b_1 c_1 \nno  - c_1 T^{\rm mat} - {5\over 2} c_1'' \ .
\een{bvsupercurrent}

Note that the transverse supercurrent $\hat{G}\uu\perp$ appears
nowhere in $G$, when written in terms of the $b_1$ and $c_1$ \zombiesns.
This is because the supersymmetry of the entire theory is now nonlinearly realized.
The field $c\ll 1$ still transforms as
a goldstone fermion of the nonlinearly realized supersymmetry:
\bbb
\{Q, c\ll 1\} = 1 + i c_1 \pp_+ c_1  \ ,
\eee
and all Virasoro primary 
operators $\co$ made out of the fields $\psi\uu {Z\uu i},~
Z\uu M$ transform as
\bbb
[Q,\co] = i c_1 \pp_+ \co \ 
\eee
(with an anticommutator replacing the commutator when $\co$ is fermionic).

\section{Physical interpretation of the $X\uu + \to \infty$ limit}
We have established that the $X\uu +\to \infty$ limit of our solution is described by 
a free worldsheet with a $bc$ ghost system 
of weights $(3/2, -1/2)$,
$D$ free scalars $Z\uu M$ and $D - 2$ 
free fermions $\psi\uu{Z\uu i}$ (as well as the
corresponding left-moving counterparts $\tilde{\psi}\uu{Z\uu i},~
\tilde{b}\ll 1,~ \tilde{c}\ll 1$ of the fermions).
The scalars have a flat metric $G\ll{MN} = \eta\ll{MN}$,
and a dilaton gradient $\pp\ll M \hat{\Phi} = V\ll M + \Delta V\ll M = \hat{V}\ll M$,
with $V\ll \m~,\Delta V\ll \m$ given in Eqns.~(\ref{vdef}, \ref{vhatdef}),
and $V\ll i = \Delta V\ll i = 0$.
The total central charge of the $Z\uu M,~\psi\uu{Z\uu i}$ system
is $26$, and the contribution of $-11$ from the
$b\ll 1 c\ll 1$ system brings the total central charge
to $15$.  In other words, the theory has critical 
central charge for a SCFT interpreted
as the worldsheet theory of an RNS superstring in conformal gauge.

\subsection{The Berkovits-Vafa construction}
This type of superconformal field theory
belongs to a class of constructions introduced in \cite{bv}.
The point of Ref.~\cite{bv} was to demonstrate
that any bosonic string theory can be cast 
as a special solution to superstring theory.  
Specifically, given any conformal field theory $T\uu{\rm mat}$ with a
central charge of $26$, it is possible to construct a
corresponding superconformal field theory 
$G ,~T $ of central charge $15$.
Upon treating the superconformal theory
as a superstring theory, the resulting physical states
and scattering amplitudes are identical to those
of the ordinary conformal theory $T\uu{\rm mat}$
when treated as a bosonic string theory.
In other words, one constructs a superconformal theory
$G ,~\tfif $ whose super-Virasoro
primaries of weight $+\hh$ naturally correspond
to the ordinary Virasoro primaries of weight unity
in the theory defined by $T\uu{\rm mat}$.\footnote{This, in turn,
is the statement that the BRST cohomologies
of the superstring theory and the corresponding bosonic theory 
are equivalent.  We have restricted ourselves to the language of old covariant
quantization rather than BRST quantization  
to avoid confusing the presentation with the introduction of
reparametrization and local supersymmetry ghosts, $\bfp,~\cfp,~\b,~\g$.  We will address the
BRST qualtization of the Berkovits-Vafa system directly in Appendix \ref{BRST}.}

The construction in \cite{bv} can be summarized as follows.
Given a conformal stress tensor $T\uu{\rm mat}$
with central charge $26$, a ghost system
$b\ll 1 c\ll 1$ can be introduced with weights $(3/2, -1/2)$ and stress
tensor $T^{b_1 c_1}$ taking precisely the form appearing in Eqn.~\rr{tbcstress} above. 
This gives rise to a fermionic primary current of
weight $3/2$:
\bbb
G \equiv  b_1 + i \nno c_1' b_1 c_1\nno - c_1 T^{\rm mat}
	- \frac{5}{2} c_1''  \ ,
\eee
which closes on the stress tensor of the theory:
\bbb
G (\s) G (\t)  \simeq  \frac{10 i}{(\t^+ - \s^+)^3} + \frac{2i}{(\t^+ - \s^+) }\tfif(\t) \ .
\eee
Here we have written the total stress tensor of the theory as
\bbb
\tfif  \equiv T\uu{\rm mat} + T\uu{b_1 c_1} \ .
\eee
This defines a superconformal theory of
central charge $15$.  To construct physical states
of the corresponding superstring theory, one starts with
a Virasoro primary state $\kket{{\cal U}}$
of weight $1$ in the theory defined by $T\uu{\rm mat}$:
\bbb
L\ll n\uu{\rm mat} \kket{{\cal U}} &=& 0\ , \qquad n\geq 1\ ,
\mmm
L\ll 0\uu{\rm mat} \kket{{\cal U}} &=& 1\ .
\label{sixpointfour}
\eee

\subsection{States and operators of the $\ct\to \infty$ limit}

Written in terms of modes, the
commutators of the $b\ll 1 c\ll 1$
theory are
\bbb
\{b\ll 1 \uu r, c\ll 1 \uu s \} = \d\ll{r,-s}\ .
\eee
In the NS sector, the super-Virasoro generators appear as
\bbb
L\ll m &=&
L\ll m\uu{\bonecone} +
L\ll m\uu{\rm mat}\ ,
\mmm
L\ll m\uu{\bonecone} &=&   
\sum\ll r
\lrdd {{3 m }\over 2} - r \rrdd
\nno b\ll 1\uu r c\ll 1\uu{m - r } \nno
+ \hh  \sum\ll r m\sqd (r - {m\over 2}) c\ll 1\uu r c\ll 1
\uu {m - r} - \hh \d\ll{m,0}\ ,
\mmm
G\ll r &=& b\ll 1\uu r + 
\sum\ll m
c\ll 1\uu{r - m} L\uu{\rm mat}\ll m
- \sum\ll{t,u} u\cc \nno c\ll 1\uu u b\ll 1\uu 
t c\ll 1 \uu {r - u - t} \nno
+  \cc\lrdd {{5r \sqd}\over 2} - {9\over 8} \rrdd c\ll 1\uu r \ .
\eee
The symbol $\nno\cc\nno$ is defined by ordering all the 
positively-moded operators to the right of negatively-moded operators.  
In the R sector, $c\ll 0$ is counted as a `positively-moded' operator, in that we
order it to the right of $b\ll 0$.\footnote{This normal ordering 
prescription is the creation-annihilation normal ordering corresponding 
to the local normal ordering $\nno\cc\nno$ used thus far.
Technically the two are distinct, though they coincide on a cylinder of infinite
radius.} 
In the R sector, the super-Virasoro generators appear as
\bbb
L\ll m &=&
L\ll m\uu{\bonecone} +
L\ll m\uu{\rm mat}\ ,
\mmm
L\ll m\uu{\bonecone} &=&   
\sum\ll r
\lrdd {{3 m }\over 2} - r \rrdd
\nno b\ll 1\uu r c\ll 1\uu{m - r } \nno
+ \hh  \sum\ll r m\sqd (r - {m\over 2}) c\ll 1\uu r c\ll 1
\uu {m - r} - {3\over 8} \d\ll{m,0}\ ,
\mmm
G\ll r &=& b\ll 1\uu r + 
\sum\ll m
c\ll 1\uu{r - m} L\uu{\rm mat}\ll m
- \sum\ll{t,u} u\cc \nno c\ll 1\uu u b\ll 1\uu 
t c\ll 1 \uu {r - u - t} \nno
+  \cc\lrdd {{5r \sqd}\over 2} + {r\over 2} - 1 \rrdd c\ll 1\uu r \ .
\eee

The normal ordering terms 
are fixed by the requirement that the 
Virasoro generators satisfy the algebra
\bbb
[L\uu{\bonecone}\ll m, L\uu{\bonecone}\ll n] = (m - n) L\ll{m + n}\uu{\bonecone}
- {{11}\over {12}} \d\ll{m,-n} (m\uu 3 - m)\ ,
\eee
and that the super-Virasoro generators satisfy 
\bbb
[L\uu{\rm mat}\ll m + L\uu{\bonecone}\ll m\cc ,\cc G\ll r]
&=& {{m - 2 r}\over 2} G\ll{m + r}\ ,
\mmm
\{ G\ll r, G\ll s\} &=& 2 
\lrdd
L\uu{\bonecone}\ll{r + s} + L\uu{\rm mat}\ll{r + s }\rrdd
+ {{c\uu{\rm tot}}\over {12}} (4r\sqd - 1)\ ,
\eee
where
\be
c\uu{\rm tot} = c\uu{\bonecone} + c\uu{\rm mat} = -11 + 26 = 15\ .
\ee  
If $c\uu{\rm mat}$ is not equal to 26, 
the super-Virasoro generators must be supplemented
with a term trilinear in the \zombie field $c_1$ to close
on the Virasoro generators.

The weight of the NS vacuum $\kket 0 \ll{b\ll 1 c\ll 1}$
is $-\hh$, so the state
\bbb
\kket{\tilde{{\cal U}}} \equiv
\kket 0 \ll{b\ll 1 c\ll 1} \ot \kket{{\cal U}}
\label{sixeleven}
\eee
has weight $+\hh$, for $\kket{{\cal U}}$ satisfying (\ref{sixpointfour}).  
It is also clear that $\kket 0 \ll{b\ll 1c\ll 1}$ is primary, 
so the product state is a Virasoro
primary of weight $\hh$ in the superconformal theory.
Since the ground state $\kket 0\ll{b\ll 1 c\ll 1}$
is annihilated by all $c\ll r$ for $r \geq 0$, 
it follows that $\kket{\tilde{{\cal U}}}$ is
annihilated by all $G\ll r$ for $r\geq \hh$.
Thus, $\kket{\tilde{{\cal U}}}$ is a super-Virasoro
primary of weight $\hh$.
In reference \cite{bv} it is shown that all physical 
states are represented (modulo null states) by 
product states of the form (\ref{sixeleven}).  
The authors also demonstrate
that all scattering amplitudes and vacuum diagrams
of the superstring theory defined by $G ,~T $
are equal to those of the bosonic theory defined
by $T\uu{\rm mat}$.   We will demonstrate this explicitly below 
using BRST methods.

It turns out that the $X\uu + \to \infty$ limit of the
\wsbb
is {\it precisely} of the form described in \cite{bv} and reviewed above, with
\bbb
T\uu{\rm mat} \equiv - {1\over\apr} \cc \eta\ll{\m\n}
\nno \pp\ll + Z\uu\m \pp\ll + Z\uu\n \nno + 
(V\ll\m + \Delta V\ll\m) \pp\ll + \sqd Z\uu\m +
{i\over 2} \sum\ll{j = 1}\uu{D - 2} :\psi\uu {Z\uu i}
\pp\ll + \psi\uu {Z\uu i}:\ .
\eee  
Our background describes a dynamical transition between 
a type 0 superstring theory in $D$ dimensions, and a bosonic
string theory in $D$ noncompact
dimensions, together with $\hh (D - 2)$
units of central charge.  The latter is generated by a current algebra
$SO(D - 2)\ll {\rm L}\ot SO(D - 2)\ll {\rm R}$ at
level 1, whose degrees of freedom are the free fermions
that used to be the superpartners of the $Z\uu i$
bosons.

\subsection{Other issues}

\heading{GSO projections}

We emphasize that our model is a solution to type 0 string
theory only, and that no direct analog for type II exists.
Our solution depends upon the existence of the type 0
tachyon, which is eliminated by the type II
GSO projection.  To put it another way, the worldsheet action
of our model explicitly breaks a discrete symmetry, left-moving
worldsheet fermion number mod 2, which is gauged in the type II string.
Furthermore, the GSO projection of the type II theory
is anomaly-free only in $10+16k$ dimensions.\footnote{In $2 + 16k$ dimensions, 
there exists a
related but different anomaly-free chiral GSO projection \cite{seibergcircles} 
with spacetime fermions in its spectrum.}
Nonetheless, the type II string, when it exists, is
closely related to the type 0 string: they are T-dual to one another on
a circle with twisted boundary conditions \cite{tdualitybetweentypeiiandtype0}.  
So we have shown
that the type II theory can be connected continuously to
bosonic string theory by combining
compactification, motion along moduli
space, and tachyon condensation.
It would be interesting to see
whether one could reach the bosonic string vacuum directly by
tachyon condensation starting from
type II on a Scherk-Schwarz circle. 

\heading{Symmetries}

Both the spacetime and internal symmetry structures are quite
different in the far past and far future.  
In the far past, the spacetime invariance is given by the $SO(D-1)$
little group of the dilaton gradient $V\ll\m$, and in the far
future the spacetime symmetry is the $SO(D-1)$ little group
of the renormalized dilaton gradient $\hat{V}\ll\m$.  There is no
natural identification between these groups.  
At intermediate times, the spacetime symmetry is broken to
the $SO(D-2)$ simultaneous little group of the dilaton
and tachyon.

Even the $SO(D-2)$ generators at early
and late times should not be identified with one another.  The
rotation generators at early times involve the $X\uu i,~ \psi\uu i$
and $\pst\uu i$ fields, whereas the rotation generators at
late times involve only the $Z\uu i$ fields.  The $M\to \infty$ 
limit of the rotation generators of the \uv theory define diagonal
$SO(D-2)$ rotations in spacetime as well as in the current algebra.
The $SO(D-2)$ rotations of $Z\uu i$ alone do not correspond
to any symmetry in the \uv theory, because finite-$M$ corrections
break $SO(D-2)\ll{Z} \otimes SO(D - 2)\ll{\rm L} \otimes SO(D-2)\ll{\rm R}$
down to the diagonal $SO(D-2)$, which can be identified with rotations in the
\uv theory.

In other words, there is
extra symmetry in the limit $M\to \infty$.
This is roughly similar to spin-orbit decoupling for heavy fermions
in nonrelativistic quantum mechanics.  
For finite masses, the little group
of the fermions' rest frame is a diagonal $SU(2)\subset SO(3,1)$ 
living in the full Lorentz group.  In the limit of infinitely heavy fermions,
the $SU(2)$ that acts on spin degrees of freedom only emerges as
an independent symmetry, conserved separately from the $SU(2)$ acting
on the spatial coordinates.  For finite fermion mass, there are small
corrections breaking the symmetry down to an overall $SU(2)$, with
coefficients that scale like $E\ll{\rm typical} / \lrdd m\ll{\rm fermion}\cc c\sqd\rrdd$. 

The appearance of enhanced gauge symmetry at late times
is intriguing.  It may be interesting to understand
the description of the unhiggsing of $SO(D-2)\ll{\rm L} \otimes SO(D-2)\ll{\rm R}$
at the level of effective field theory.

%
%
%
Interesting effects arise from broken Lorentz invariance in cosmological
backgrounds such as the one studied here.
At finite $X^+$, the target-space 
gauge symmetry is broken spontaneously
by a field transforming in the $(D-2,D-2)$ of the
gauge group.    This field decays exponentially to zero at late times.
Rather than a scalar, the field is a rank-two tensor under the Lorentz group, 
and has $m\sqd =  2/\apr$.
%
%
%

\section{Discussion and conclusions}
\begin{figure}[htb]
\begin{center}
\includegraphics[width=3.2in,height=3.2in,angle=0]{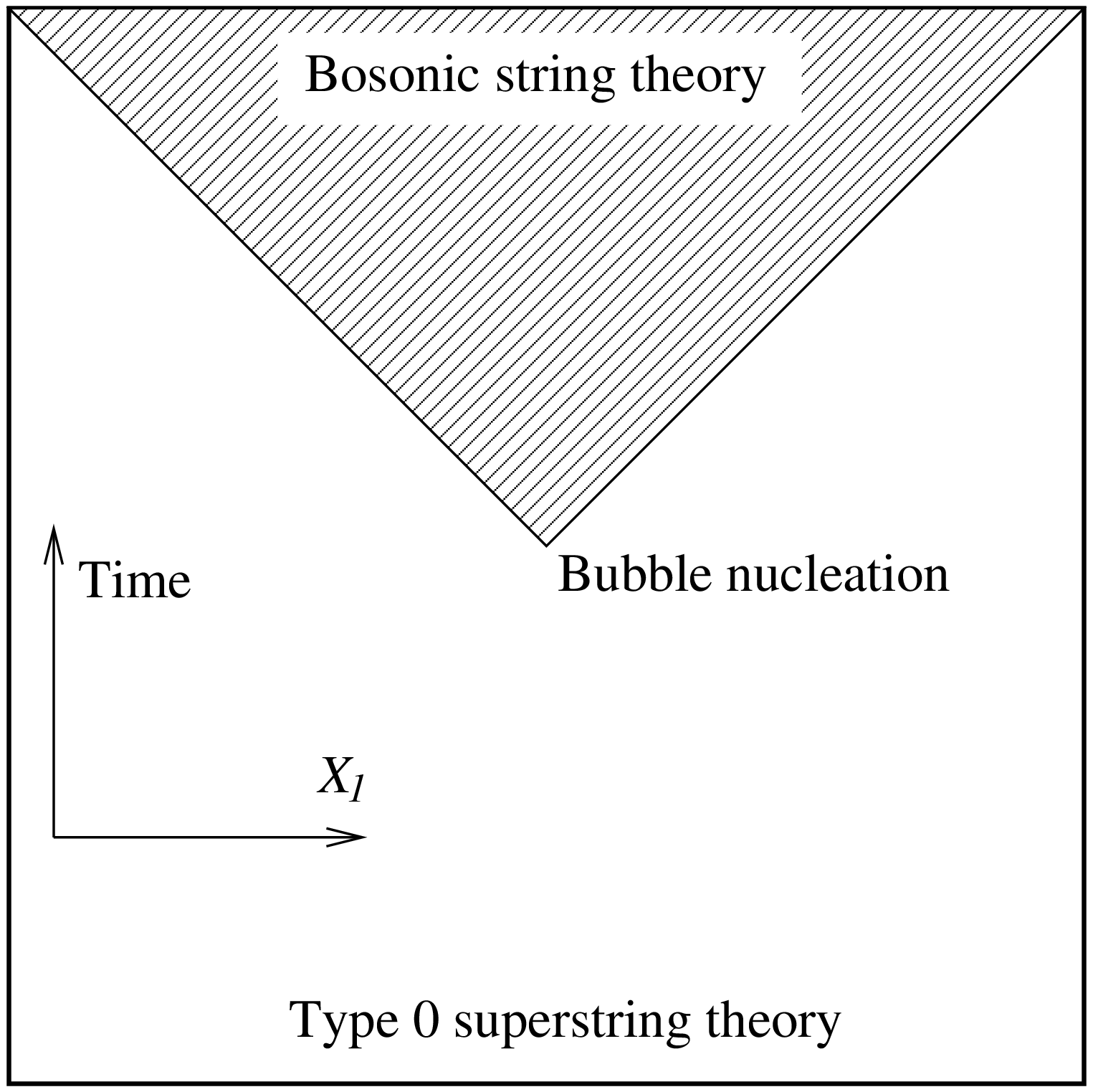}
\caption{The phase transition from type 0 superstring theory 
to bosonic string theory.
After nucleation of the bubble, the solution interpolates in
time between the two theories. Our solution focuses on the upper left-hand
corner of the diagram, where the bubble is approximated by a domain wall moving to
the left at the speed of light.}
\label{bubblefig}
\end{center}
\end{figure}

In this paper we have introduced exact solutions describing
dynamical transitions among string theories that
differ from one another in their worldsheet
gauge algebra.  The transitions follow an instability in an initial $D$-dimensional 
type 0 theory.  The dynamics spontaneously break worldsheet supersymmetry, 
giving rise to a bosonic string theory in the same number of dimensions
deep inside the tachyonic phase.  The final theory exhibits 
$SO(D-2)\ll{\rm L} \otimes SO(D-2)\ll{\rm R}$ gauge symmetry carried 
by current algebra degrees of freedom.  This transition is depicted schematically
in Fig.~\ref{bubblefig}.

\ \\

Though our initial state is type 0 string theory in $D>10$, one can begin instead from
a state with lightlike linear dilaton in the critical dimension, or spacelike linear dilaton 
in subcritical dimensions.  
If we have $D = 10$ with $\Phi = - {q\over{\sqrt{2}}} \cc X\uu -$, or $2 \leq  D < 10$
with $\Phi = q X\ll 1$ and $q = \sqrt{{D - 10}\over {4\apr}}$,
we can consider the tachyon profile $\ct = \m\cc\exp{\b X\uu +}$.  
The physical interpretation may be different from that in $D>10$,
but the CFT is equally solvable for all values of $D\geq 2$.
For $D=2$, time-dependent tachyon solutions have been
studied using the matrix model \cite{mm1, mm2, mm3, mm4}.
Our model is similar to the example of \cite{mm2}, distinguished
by our initial state (which is type 0 rather than bosonic
string theory, and without a Liouville wall).   It would be interesting to study 
the transition described in this paper using the technology
of the matrix model.

We have given a conclusive answer to the question of
whether bosonic string theory can be connected
with superstring theory by a dynamical
realization of the formal embedding of \cite{bv}.
This possibility has been anticipated (see,
for example, \cite{joelectures}).  The surprising feature 
in our model is the essential role of time dependence.
It was speculated in \cite{joelectures} that the bosonic
string may appear as a stationary but non-static solution
of the superstring; the simple scenario we present here
realizes bosonic string theory as a type 0 background that 
is not even stationary.

The achievement of \cite{bv} was to prove that
all string vacua, even those with distinct worldsheet 
gauge algebras, can be unified into a single
`universal string theory.'\footnote{Extensions of this work include
\cite{Ohta:1994qp,Bastianelli:1994cb,Bastianelli:1994xb}.}    
For the vacua of bosonic and
type 0 strings, we have promoted this formal unification
to a physical one, constructing exact, time-dependent solutions that 
change the effective gauge algebra of the worldsheet.  In this
context, the moral of our result is clear:  
{\it The universal string theory
is a cosmology. } We expect exploration of
string dynamics in $D>10$ to yield further surprises.

\section*{Acknowledgments}
We thank Ofer Aharony, Mark Jackson, Juan Maldacena, Joseph Polchinski,
Savdeep Sethi, Eva Silverstein and Edward Witten for valuable conversations.
We also thank David Kutasov and Ilarion Melnikov for additional useful discussions.
S.H.~is the D.~E.~Shaw \& Co.,~L.~P.~Member
at the Institute for Advanced Study.
S.H.~is also supported by U.S.~Department of Energy grant DE-FG02-90ER40542. 
I.S.~is the Marvin L.~Goldberger Member
at the Institute for Advanced Study, and is supported additionally
by U.S.~National Science Foundation grant PHY-0503584. 

\setcounter{equation}{0}
\numberwithin{equation}{section}


%
%

\appendix{Generating the canonical transformation from UV to IR fields}
The transformation in Eqns.~\rr{firstcanonical}, \rr{xshift}
mixes fields in a nonlinear way,
and it is not immediately obvious that the transformation preserves
commutators.  A simple way to check that a finite change of variables
is canonical is to derive the transformation rules from
a \it generating function of the first kind\rm\cite{goldstein}.  
To see how this works in the case at hand we move to canonical variables,
defining conjugate momenta as derivatives of the action:
\bbb
\begin{array}{cc}
{\bf q}\uu + \equiv X\uu +\ , &
{\bf q}\ll - \equiv \delta_{X^-}{\cal L} = - {1\over{2\pi\apr}} \dot{X}\uu +\ ,
\\ & \\
{\bf Q}\uu - \equiv \delta_{Y\uu +}{\cal L} = - {1\over{2\pi\apr}} \dot{Y}\uu -\ ,
& 
{\bf Q}\ll + \equiv Y\uu -\ .
\end{array}
\eee
This translates into the following conjugate momenta:
\bbb
\begin{array}{cc}
{\bf p}\ll + = \delta\ll{X\uu +} {\cal L} = - {1\over{2\pi\apr}} \dot{X}\uu -\ ,
&
\quad {\bf p}\uu - = - X\uu -\ , \quad
\\ & \\
{\bf P}\ll - = - Y\uu +\ ,
& 
{\bf P}\uu + = \delta\ll{Y\uu -}{\cal L}\ .
\end{array}
\eee
We also obtain fermionic canonical coordinates, and momenta as derivatives of the action
acting from the left:
\bbb
\begin{array}{cc}
{\bf q}\equiv \psi\uu +\ ,
&
\tilde{{\bf q}} \equiv \pst\uu +\ ,
\\ & \\
{\bf Q}\equiv c\ll 4\ ,
&
\tilde{{\bf Q}}\equiv \tilde{c}\ll 4\ ,
\\ & \\
~~ {\bf p} = +{ i\over{2\pi}} \psi\uu -\ , ~~
& 
~~ \tilde{{\bf p}} = + {i\over{2\pi}}  \pst\uu -\ ,  ~~
\\ & \\
{\bf P} = - {i\over{2\pi}}  b\ ,
&
\tilde{{\bf P}} =  - {i\over{2\pi}}     \tilde{b}\ .
\end{array}
\eee

At this point we can define our canonical transformation using a
generating function $F\ll 1({\bf q},{\bf Q})$, such that
\bbb
{\bf p}\ll i \equiv + {{\d F\ll 1}\over{\d {\bf q}\uu i}}\ ,
& \qquad  &
{\bf P}\ll a \equiv - {{\d F\ll 1}\over{\d {\bf Q}\uu a}}\ ,
\mmm
\tilde {\bf p} \equiv + {{\d F\ll 1}\over{\d \tilde {\bf q}}}\ ,
& \qquad  &
\tilde {\bf P} \equiv - {{\d F\ll 1}\over{\d \tilde {\bf Q}}}\ ,
\mmm
 {\bf p} \equiv + {{\d F\ll 1}\over{\d  {\bf q}}}\ ,
& \qquad  &
{\bf P} \equiv - {{\d F\ll 1}\over{\d  {\bf Q}}}\ ,
\een{genfunrules}
where all fermionic derivatives are again understood to act
from the left.   Specifically, our generating function takes the following 
form
\bbb
F\ll 1 & \equiv & \int d \s\ll 1 {\cal F}\ll 1\ ,
\nn\\ && \nn\\
{\cal F}\ll 1 &\equiv &
-\frac{1}{2\pi\apr} \left( X^+ \dot Y^- - \dot X^+ Y^- \right)
\nn\\ && \nn\\ &&
- {i\mu \over{2\pi}}\exp{\b X^+}  
\biggl[~
 \pst\uu + c\ll 4 -  \psi\uu + \tilde{c}\ll 4
 -  (c\ll 4 \tilde{c}\ll 4\pr - c\ll 4\pr \tilde{c}\ll 4)
-  \b\, \dot X^+ c\ll 4\tilde{c}\ll 4
~\biggr]
\nn\\ && \nn\\
& = &
{\bf q}\uu + {\bf Q}\uu - - {\bf q}\ll - {\bf Q}\ll + 
\nn\\ && \nn\\ &&
- {i\mu \over{2\pi}} \exp{\b {\bf q}^+}
\biggl[\,
{\bf \tilde q Q} - {\bf q \tilde Q}
- ( {\bf Q \tilde Q' - Q' \tilde Q})
+2 \pi \alpha' \b\, {\bf q}_- {\bf Q \tilde Q}
\, \biggr]\ .
\eee
Following the rules in Eqn.~\rr{genfunrules},
we obtain eight transformation equations, one for each of the fields
$\psi^+,~ \psi^-,~ \tilde\psi^+,~ \tilde\psi^-,~b,~ \tilde b,~Y^+,~Y^-$. 
These comprise a system of equations that can be solved to express 
all of the lower-case canonical variables in terms of the upper-case 
canonical variables, or vice-versa.  For the $\psi$ fields, we obtain
\bbb
\psi \uu + &=&  2 c^\prime \ll 4 - M\uu{-1} \tilde{b}\ll 4
 + 2\b (\pp\ll + X\uu +) c\ll 4 \ ,
\nn \\ && \nn \\
\psi\uu - &=& M \tilde{c} \ll 4\ ,
\nn \\ && \nn \\
\pst\uu + &=&  - 2  \tilde{c}^\prime \ll 4 + M\uu {-1} b \ll 4 
+ 2 \b (\pp\ll - X\uu +) \tilde{c}\ll 4 \ ,
\nn \\ && \nn \\
\pst\uu - &=& - M c\ll 4\ .
\eee
Similarly, the $X$ fields transform according to
\bbb
X\uu + &\equiv& Y\uu +\ ,
\mmm
X\uu - &\equiv& Y \uu - + i \b\apr \m\cc\exp{\b X\uu +} 
	c\ll 4 \tilde{c}\ll 4 \ .
\eee
Finally, the $b,~\tilde b$ fields obey
\bbb
b_4 & = & -2\b \pp_+ X^+ \psi^- + 2 {\psi'}^- + M \tilde \psi^+\ ,
\mmm
\tilde b_4 & = & 2\b \pp_+ X^+ \psi^- - 2 \tilde {\psi'}^- - M\psi^+\ .
\eee
These are all consistent with the field redefinitions used in Section 
\ref{canonical} above, and we conclude that they are proper, canonical transformations.

We are left with a stress tensor and
supercurrent describing two free bosons
$Y\uu\pm$ and a free $bc$ ghost system
with weights $(3/2, -1/2)$.  
The new equal-time commutators are
\bbb
\{b\ll 4 (\s), c\ll 4(\t)\} &=& + 2\pi \d(\s\ll 1 - \t\ll 1)\ ,
\mmm
\{\tilde{b}\ll 4 (\s) , \tilde{c}\ll 4(\t) \} &=& + 2 \pi \d(\s\ll 1 - \t\ll 1)\ ,
\mmm
\left[\pp\ll\pm Y\uu\m(\s), \pp\ll\pm Y\uu\n(\t)\right] &=& - \pi i \apr
\eta\uu{\m\n} \d\pr(\s\ll 1 - \t\ll 1)\ ,
\eee
and the rest vanish.

\appendix{OPEs in the interacting $2D$ theory}
In this appendix we record a number of OPEs that
are needed to compute quantum corrections
to the transformations of the stress tensor and
supercurrent under the transformation
\rr{lorentzinvar0}.

\subsection{Solving for the OPE of $\psi\uu -$ with $\pst\uu -$}
To compute the quantum contribution to the
central charge, we will need to give a sensible,
consistent definition to the 
composite operators that enter the
stress tensor and supercurrent.  We define the
composite operators of the $X\uu\m, \psi\uu\m, 
\pst\uu\m$ theory in terms of the $:\cc\cc:$
normal ordering prescription, whose properties
we list in Section \ref{normalorderingdefinitions}.
However, we have to choose more specific definitions
of composite operators involving the fields 
$Y\uu\m, b\ll 4, c\ll 4, \tilde{b}\ll 4, \tilde{c}\ll 4$,
as well as $X\uu\m, \psi\uu\m, \pst\uu\m$.

We begin by solving for the $\psi\uu -(\s) \pst\uu - (\t)$ OPE.
To do this, we must rely on the normal ordering prescription 
for $\psi\uu \pm$.  Based on diagrammatic arguments, we see that
the OPEs for fields involving a $\psi\uu +$ are unaffected by 
the interaction terms.
So the OPEs for $\psi\uu \mp$ with $\psi\uu\pm$ 
and $\pst\uu\pm$ with $\pst\uu\mp$ are
\bbb
\psi\uu \mp (\s) \psi\uu \pm (\t) =
:   \psi\uu \mp (\s) \psi\uu \pm (\t) :
+ {i \over{\t\uu + - \s\uu +}}\ ,
\een{firstope}
\bbb
\pst\uu\mp (\s) \pst\uu\pm (\t) =
:   \pst\uu \mp (\s) \pst\uu \pm (\t) :
+ {i \over{\t\uu - - \s\uu -}}\ .
\een{secondope}
By acting on Eqn.~\rr{firstope} with $- \hh M (\t\uu + , \t\uu -)$ and using the equation of
motion $\pp\ll + \pst\uu - = - \hh M \psi\uu +$, 
we can find a first-order partial differential equation
for the OPE of $\psi\uu -$ with $\pst\uu -$:
\bbb
\pp\ll {\t\uu +}\lrdd \psi \uu - (\s) \pst\uu - (\t) \rrdd
= 
\pp\ll {\t\uu +}
\lrdd : \psi \uu - (\s) \pst\uu - (\t) : \rrdd
- {i\over 2} {{M(\t\uu + , \t\uu -)}\over{\t\uu + - \s\uu +}}\ .
\een{firstequation}
We can also act on Eqn.~\rr{secondope} with $+ \hh M (\s\uu + , \s\uu -)$ and
use the equation of motion $\pp\ll - \psi\uu - = + \hh M \pst\uu +$
to obtain
\bbb
\pp\ll {\s\uu -}\lrdd \psi \uu - (\s) \pst\uu - (\t) \rrdd
= 
\pp\ll {\s\uu -}
\lrdd : \psi \uu - (\s) \pst\uu - (\t) : \rrdd
+ {i\over 2} {{M(\s\uu + , \s\uu -)}\over{\t\uu - - \s\uu -}}\ .
\een{secondequation}


Any two simultaneous solutions to Eqns.~\rr{firstequation} and \rr{secondequation} must
differ by a bilocal operator depending only on $\s\uu +$ and $\t\uu -$. 
We would like to find the most general possible OPE that
satisfies both the constraints of conformal invariance and equations of
motion (\ref{firstequation}, \ref{secondequation}).  
The basic procedure can be divided into three steps.  
First, we find a particular OPE satisfying the equation
of motion at the level of terms
that are singular as $\s$ approaches $\t$.  
Then, we add a set of terms with smooth dependence on $\s - \t$, 
so that the equations of motion (\ref{firstequation}, \ref{secondequation})
are satisfied identically.  
Finally, we construct the full set of
possible solutions to the equations (\ref{firstequation}, \ref{secondequation}).
Since the equations define a linear, inhomogeneous system of partial differential 
equations with operator-valued source term,  
we can start by constructing the general solution to
the corresponding system of {\it homogeneous} equations.  By adding this to
our particular solution we find the general solution to (\ref{firstequation}, 
\ref{secondequation}).

\heading{1: Constructing a particular solution for the singular terms}
First, we construct a particular simultaneous 
solution that satisfies Eqns.~(\ref{firstequation}, \ref{secondequation})
up to terms that have an infinite number of continuous derivatives as
$\s\to\t$.
The singular terms in the OPE sum up to
\bbb
  \psi \uu - (\s) \pst\uu - (\t) &\simeq&  : \psi \uu - (\s) \pst\uu - (\t) :
 - { i \over 2} \ln {1\over{L\sqd}}\lba (\t\uu + - \s\uu +) (\t\uu - -
\s\uu -) \rba M(\s\uu +, \t\uu -) 
\mmm
&&
+ \lrdd {\rm smooth~as~} \s\to \t \rrdd\ ,
\een{singulartermsintheope}
where $L$ is an arbitrary length scale.
It is straightforward to check that the
OPE in Eqn.~\rr{singulartermsintheope} satisfies Eqns.~\rr{firstequation} 
and \rr{secondequation},
up to terms smooth as $\s\to\t$.

\begin{figure}[htb]
\begin{center}
\includegraphics[width=4.1in,height=2.5in,angle=0]{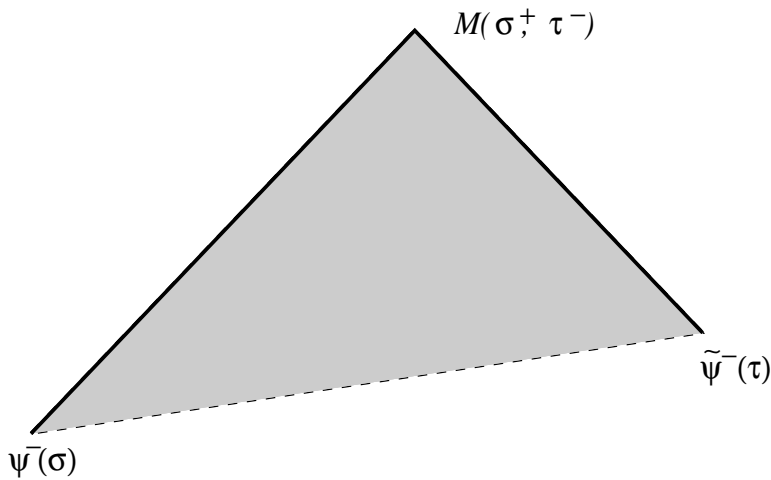}
\caption{Light rays emerge from the insertion points of $\psi\uu -(\s)$ and
$\tilde \psi\uu - (\tau)$.  The operator $M \equiv \exp{\b X^+}$ on the right-hand side
of the OPE is evaluated at the intersection point $(I)\equiv (\s\uu +, \t\uu -)$ 
of the light rays.  The coefficient function 
$\left(\ln \left|\frac{1}{L^2}(\t^+ - \s^+)(\t^- - \s^-)\right| \right)$
is the logarithm of the area of the triangle formed by the light rays and the 
line joining the insertion points. }
\label{orderlegs}
\end{center}
\end{figure}

The singular term in the OPE has a simple representation in
terms of the geometry of the Lorentzian worldsheet.  Drawing two
intersecting light rays from the points of insertion of the operators $\psi\uu -,~\pst\uu -$,
one finds that the two light rays, together with the line joining the points, define a 
triangle.   The logarithm of the area of this triangle is the singular
coefficient function in the OPE.  The operator being multiplied by this
coefficient function is evaluated at the intersection point of the two light rays.
This framework is depicted in Fig.~\ref{orderlegs}.  
This very simple Lorentzian description of the OPE suggests that the Lorentzian
worldsheet is the natural setting for the type of CFT of interest,
rather than the $2D$ Euclidean continuation.

\heading{2: Adding smooth terms to satisfy the equations of motion}
We can solve for the smooth terms in the OPE as well.
One solution to Eqns.~\rr{firstequation} and \rr{secondequation} takes the form
\bbb
\psi \uu - (\s) \pst\uu - (\t) &=& : \psi \uu - (\s) \pst\uu - (\t) :
- { i \over 2} \ln \lba {1\over{L\sqd}} (\t\uu + - \s\uu +) (\t\uu - -
\s\uu -) \rba M(\s\uu +, \t\uu -) 
\nn\\ && \nn\\
&&
- {i\over 2} \sum\ll{n = 1}\uu\infty {1\over{n\cdot n!}} \lsqq
(\t\uu + - \s\uu +)\uu n \cc \pp\ll +\uu n + (\s\uu - - \t\uu -)\uu n \cc
\pp\ll - \uu n \rsqq
M(\s\uu + , \t\uu -)\ .
\een{totalminusminusopeapp}
The OPE \rr{totalminusminusopeapp} identically satisfies the equations of motion,
including terms that are smooth as $\s$ approaches $\t$.


\heading{3: Constructing the general solution}
Any other solution must still satisfy Eqns.~\rr{firstequation} and \rr{secondequation},
and therefore must differ from the right-hand side of Eqn.~\rr{totalminusminusopeapp} by the
addition of an operator $\Delta \co(\s, \t)$, satisfying
\be
\pp\ll {\s\uu -} \Delta \co(\s,\t) = \pp\ll{\t\uu +} \Delta \co(\s,\t) = 0\ .
\label{bseven}
\ee
We are therefore lead to consider a completely
general OPE of two operators, one at $\s$ and
the other at $\t$.  
Any operator on the right-hand side of the OPE can be expressed in the form
\bbb
\sum\ll A {\cal C}\ll{(A)}\uu{{\bf 2}} (\s - \t) \co\uu{(A)} (\t\uu +, \t\uu -)\ ,
\eee
where the expansion is performed around the location $\t$ of the second
operator on the left-hand side.
The right-hand side of the same OPE can be expressed as 
\bbb
\sum\ll A {\cal C}\ll{(A)}\uu{\rm {\bf 1}} (\s - \t) \co\uu{(A)} (\s\uu +, \s\uu -)\ ,
\eee
with the expansion around the location $\s$ of the first operator
on the left-hand side.  Since the coefficient functions 
${\cal C}\ll{(A)}\uu{{\bf 1}} (\s - \t)$
can be rewritten in terms of the coefficient functions
${\cal C}\ll{(A)}\uu{\bf 2} (\s - \t)$,
we can expand a given OPE around whichever point is most convenient.
Of course, we can equally well expand the OPE around the midpoint between the two operators,
or around any other point in their vicinity.

Now we would like to parametrize our freedom to change
the OPE \rr{totalminusminusopeapp} in a way that is 
consistent with the equations of motion. To do this, let us
expand around the point $(I)\equiv (\s\uu +, \t\uu -)$, 
where the two light rays emanating from $\s$ and $\t$ intersect (see Fig.~\ref{orderlegs}).  
We write the additional term $\Delta \co$ on the right-hand side of our OPE as
\bbb
\Delta \co (\s,\t) \equiv \sum\ll A {\cal C}\ll{(A)}
 \uu{\rm I}(\s - \t) \co\uu{(A)} (\s\uu + , \t\uu -)\ .
\label{bten}
\eee
In terms of the basis of operators at point $(I)$, the condition $\pp\ll {\s\uu -} \Delta \co
= \pp\ll{\t\uu +} \Delta \co = 0$ means that the coefficient functions 
${\cal C} \uu{\rm I}\ll{(A)}(\s - \t) $ must be constants ${\bf C}\ll{(A)} $.  The
operator $\Delta \co (\s,\t)$ is then given by
\bbb
\Delta \co (\s,\t) \equiv \sum\ll A~{\bf C}\ll{(A)}  {\cal O}\uu{(A)} (\s\uu +,\t\uu -)\ .
\eee
Rewriting this as an expansion at $\t$, we have
\bbb
\Delta \co (\s,\t) = \sum\ll A~
\sum\ll {n = 0} {{{\bf C }\ll{(A)}}\over{n!}}  (\s\uu + - \t\uu +)\uu n \pp\uu n\ll + {\cal O}
\uu{(A)}
 (\t\uu +,\t\uu -)\ .
\eee
Since the operators on the left-hand side
of the OPE are $\psi\uu - (\s)$ and $\pst\uu - (\t)$,
conformal invariance forces all the ${\cal O}\uu{(A)}$ to have
weight $(\hh, \hh)$.

We can further prove that any contribution to the right-hand side of the
OPE must scale with $M$ as $M\uu {+1}$.  The only
Feynman diagrams that can contribute to this OPE
have two untruncated lines labeled $\psi\uu -,~\pst\uu -$
representing the left-hand side
and some number of truncated lines representing the right-hand side.
There are no connected diagrams in the theory
with more than one interaction vertex.
This means that the right-hand side of the OPE scales as $M\uu {+1}$.  The only local operator
${\cal O}\uu{(A)}$ of weight $(\hh, \hh)$ that scales as
$M\uu {+1}$ is $M = \m \cc\exp{\b X\uu +}$ itself.

Since we have not specified the distance scale $L$ in the
argument of the logarithm appearing in Eqn.~\rr{totalminusminusopeapp},
the freedom to shift the right-hand side of the OPE by $M(\s\uu +, \t\uu -)$ is
already present in the freedom to choose $L$.  Thus,
Eqn.~\rr{totalminusminusopeapp} is already the most general allowed OPE of
$\psi\uu -(\s)$ with $\pst\uu - (\t)$.

\subsection{Operator product expansion of $X\uu -$ with $\psi\uu -$
and $\pst\uu -$}
\label{xminuspsiminusopesection}
We can use similar techniques to fix the OPE of $X\uu -$ with $\psi\uu -$.
The result must respect conformal invariance and the equations of motion.
The OPE must also have only $X\uu +$ and its superpartners on the right-hand side,
and the right-hand side must scale as a single power of $M\uu {+1}$.  In this case
we will only derive the singular terms.

Using the equation of motion $\pp\ll - \psi\uu - = \hh 
M\cc \pst\uu +$ and the OPE \rr{opexplusxminus},
we obtain
\bbb
&&
\pp\ll{\t\uu -} \lrdd X\uu - (\s) \psi\uu - (\t) \rrdd =
\pp\ll{\t\uu -} \cc 
\lrdd
: X\uu - (\s) \psi\uu - (\t) : \rrdd
\mmm
&&
\kern+80pt
+ {{\b \apr}\over 4} \ln\lsqq - {1\over{L\sqd}}
(\s\uu + - \t\uu + )(\s\uu - - \t\uu - ) \rsqq
M(\t) \pst\uu +(\t)\ .
\een{newequation1}
By then using the equation of motion $\pp\ll + \pp\ll - X\uu -
=   - i {{\b\apr }\over 4} \cc M \pst\uu + \psi\uu +$ and the OPE
\rr{firstope}, we find
\bbb
\pp\ll{\s\uu +} 
 \pp\ll{\s\uu -}  \lrdd X\uu - (\s) \psi\uu - (\t) \rrdd &=&
  \pp\ll{\s\uu +} 
 \pp\ll{\s\uu -}
 \cc 
\lrdd
: X\uu - (\s) \psi\uu - (\t) : \rrdd
\mmm
&&
+   {{\b\apr}\over 4}
{1\over{\t\uu + - \s\uu +}}
 M (\s\uu + , \s\uu -) \pst\uu + (\s)\ .
\een{newequation2}

We will only solve the system of 
equations \rr{newequation1}, \rr{newequation2} up to 
nonsingular terms.  One particular solution that sums
all the singular contributions exactly is
\bbb
X\uu - (\s) \psi\uu - (\t)
 &=&  : X\uu - (\s) \psi\uu - (\t):
\mmm
&&
\kern-20pt
+{{\b\apr}\over 4} \ln\lsqq - {1\over{L\sqd}}
(\s\uu + - \t\uu + )(\s\uu - - \t\uu - ) \rsqq
\int\ll{\s\uu -}\uu{\t\uu -}
dy M(\t\uu +, y) \pst\uu + (y) 
\mmm
&&
\kern-20pt
+ \lrdd{\rm smooth~as~}\s\to\t \rrdd\ . 
\eee
The right-hand side of the $X\uu - (\s) \psi\uu - (\t)$ OPE can be shifted by
a general solution $\Delta {\cal O}\ll 2 $ to the {\it homogeneous} versions
of Eqns. \rr{newequation1}, \rr{newequation2}:
\bbb
\Delta {\cal O}\ll 2 = \Delta {\cal O}\ll{2\uu +}(\t\uu +, \s\uu +)
+ \Delta {\cal O}\ll{2\uu -} (\t\uu +, \s\uu -)\ .
\eee

Let us now constrain the OPE by appealing to arguments parallel
to those starting with Eqn.~(\ref{bten}) above.  Let us first focus
on potential shifts of the right-hand side of the OPE contained in
$\Delta {\cal O}\ll{2\uu -} (\t\uu +, \s\uu -)$.  We can generically
expand in local operators around the intersection point $(I') \equiv (\t\uu +, \s\uu -)$:
\bbb
\Delta \co \ll {2\uu -} (\t\uu +,\s\uu -) \equiv 
\sum\ll A {\cal C}\ll{(A)}
 \uu{\rm I}(\s - \t) \co\uu{(A)} (\t\uu + , \s\uu -)\ .
\eee
By the same arguments below Eqn.~(\ref{bten}), the coefficient functions 
${\cal C}\ll{(A)} \uu{\rm I}(\s - \t)$ must be constants, which we again label
by ${\bf C}_{(A)}$.  Expanding at $\s$, we obtain
\be
\Delta \co \ll {2\uu -} (\t\uu +,\s\uu -) = \sum\ll A~
\sum\ll {n = 0} {{{\bf C }\ll{(A)}}\over{n!}}  (\t\uu + - \s\uu +)\uu n \pp\uu n\ll 
+ {\cal O} \uu{(A)} (\s)\ .
\ee
By conformal invariance of the OPE, the weight of 
${\cal O} \uu{(A)}$ must be $(0,\hh)$.  
However, each ${\cal O} \uu{(A)}$ must have one factor of $M^{+1}$, as well as an 
odd number of fermions, making the total weight of each ${\cal O}\uu{(A)}$
at least $\frac{3}{2}$.  We conclude that 
$\Delta \co \ll {2\uu -}(\t\uu +, \s\uu -) = 0$.

Let us now consider contributions from $\Delta {\cal O}\ll{2\uu +}(\t\uu +, \s\uu +)$.
The operator $\Delta {\cal O}\ll{2\uu +}$ is bilocal and right-moving in each argument.
We can expand $\Delta {\cal O}\ll{2\uu +}$ in right-moving local operators with coefficient
functions depending on the difference $\s\uu + - \t\uu +$.
Each term in the expansion
has exactly one factor of $M^{+1}$, which has left-moving weight $\hh$.  Any $\pp\ll -$
derivatives contained in the local operators can only increase the left-moving weight, and 
the coefficient functions in the expansion 
are independent of the ``$-$'' coordinates.  The left-hand side of the OPE $X\uu - \psi\uu -$ 
has total left-moving weight zero, so we conclude 
$\Delta {\cal O}\ll{2\uu +}(\t\uu +, \s\uu +) = 0$.

%
%

Hence, up to smooth terms, 
we have completely fixed the OPE to be
\bbb
X\uu - (\s) \psi\uu -(\t) & \simeq&
: X\uu - (\s) \psi\uu -(\t) :
\mmm
&&
\kern-35pt
+{{\b\apr}\over 4} \ln\lsqq - {1\over{L\sqd}}
(\s\uu + - \t\uu + )(\s\uu - - \t\uu - ) \rsqq
\int\ll{\s\uu -}\uu{\t\uu -}
dy M(\t\uu +, y) \pst\uu + (y) \ .
\eee
For fermions with ``$\widetilde{\phantom{A}} $'' labels, we can apply the same
reasoning to derive the OPE
\bbb
X\uu - (\s) \pst\uu -(\t) &\simeq&
: X\uu - (\s) \pst\uu -(\t) :
\mmm
&&
\kern-35pt
- {{\b\apr}\over 4} \ln\lsqq - {1\over{L\sqd}}
(\s\uu + - \t\uu + )(\s\uu - - \t\uu - ) \rsqq
\int\ll{\s\uu +}\uu{\t\uu +}
dy M(y, \t\uu -) \psi\uu + (y) \ .
\eee
As above, the notation $\simeq$ denotes equality up to
the addition of smooth terms.

\subsection{Solving for the remaining OPEs}
\begin{figure}[htb]
~~~~~~~~~~~~~~~\includegraphics[width=5.35in,height=0.38in,angle=0]{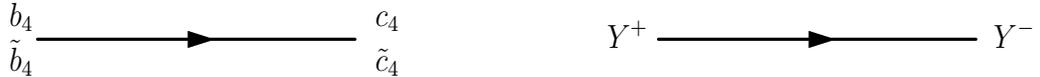}
\caption{Propagators in the $b\ll 4,~c\ll 4$ theory are oriented, 
flowing from $b\ll 4 ,~\tilde{b}\ll 4 $ 
to $c\ll 4 ,~\tilde{c}\ll 4 $ (left panel), 
and from $Y\uu +$ to $Y\uu -$ (right panel).}
\label{feynman2}
\end{figure}

\begin{figure}[htb]
~~~~~~~~~~~~\includegraphics[width=4.1in,height=1.0in,angle=0]{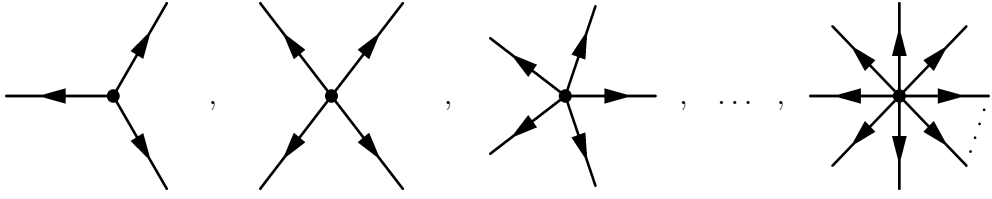}
\caption{Because the propagators in the theory are oriented, all interaction
vertices are composed strictly of outgoing lines.  This is true for both the 
\ir and \uv theories.}
\label{feynman3}
\end{figure}

We would now like to solve for the OPE between $c\ll 4$ and
$\tilde{c}\ll 4$.   We can perform a series of steps analogous to the 
derivation of the OPE between $\psi\uu -$ and $\pst\uu -$.
First, we note that all OPEs involving $b\ll 4,~\tilde{b}\ll 4$ are,
as usual,
\bbb
b\ll 4 (\s) c\ll 4 (\t) \sim
+ {i\over{\s\uu + - \t\uu +}} \ ,
\xxx
\tilde{b}\ll 4 (\s) \tilde{c}\ll 4 (\t)
\sim 
 +  {i\over{\s\uu - - \t\uu -}}\ ,
\eee
with no other singular terms.
As described in Section \ref{opes}, we 
define a normal ordering prescription $\nno\co\ll 1 \co\ll 2 \nno$
that subtracts the singularity whenever one of the operators involved is
$b\ll 4, \tilde{b}\ll 4$ or $Y\uu +$:
\bbb
b\ll 4 (\s) c\ll 4 (\t) 
 = \nno b\ll 4 (\s) c\ll 4 (\t)\nno  + {i\over{\s\uu + - \t\uu +}}\ ,
\xxx
\tilde{b}\ll 4 (\s) \tilde{c}\ll 4 (\t)
= \nno   \tilde{b}\ll 4 (\s) \tilde{c}\ll 4 (\t)  \nno 
 +  {i\over{\s\uu - - \t\uu -}}\ .
\eee
The consistency of
this prescription follows from the structure of the Feynman diagrams
in the $b\ll 4 c\ll 4$ theory, where the Lagrangian has
oriented propagators (flowing from
$b\ll 4 ,~\tilde{b}\ll 4 $ to $c\ll 4 ,~\tilde{c}\ll 4 $
and $Y\uu +$ to $Y\uu -$).  These propagators are displayed in
Fig.~\ref{feynman2}.  As a consequence, interaction vertices
have only outgoing lines, as shown in Fig.~\ref{feynman3}.  
(It should be emphasized that the absence of loop contributions to 
interaction diagrams holds for both the \ir as well as the \uv systems.)

\ \\

\ \\
\ \\

Since the operators $b\ll 4,~ \tilde{b}\ll 4$ are related to
$c\ll 4,~ \tilde{c}\ll 4$ by differential equations, we can write
a system of equations for the OPE $c_4(\s) \tilde{c}_4(\t)$, and
solve it as we did for the OPE $\psi\uu - (\s) \pst\uu -(\t)$.
Defining 
\be
N(\s) \equiv M\uu{-1} (\s) = {1\over\m} \exp{- \b X\uu +}\ ,
\ee
the resulting OPE takes the form
\bbb
 c\ll 4 (\s) \tilde{c}\ll 4 (\t) 
 &=& 
\nno c\ll 4 (\s) \tilde{c}\ll 4 (\t) \nno
- { i \over 2} \ln\lba {1\over{L\psq}} (\t\uu + - \s\uu +) (\t\uu - -
\s\uu -) \rba N(\s\uu +, \t\uu -)
\nn\\
&-& {i\over 2} \sum\ll{n = 1}\uu\infty {1\over{n\cdot n!}} \lsqq
(\t\uu + - \s\uu +)\uu n \cc \pp\ll + \uu n + (\s\uu - - \t\uu -)\uu n \cc
\pp\ll - \uu n \rsqq
N(\s\uu + , \t\uu -)\ .
\een{opecctilde}
This is the unique solution to the OPE for $c_4$ with $\tilde{c}_4$
that satisfies the equations of motion and the constraints of conformal invariance.  
The only freedom in the solution is the freedom to choose the length scale
$L\pr$.

There is also a logarithmic correction to the OPE of $Y\uu -$ with $c\ll 4 $ and
$\tilde{c}\ll 4 $, proportional to $N(\s)$.  Making use of
the same techniques employed in Section \ref{xminuspsiminusopesection},
we find:
\bbb
Y\uu - (\s) c\ll 4 (\t) &\simeq&
\nno Y\uu - (\s) c\ll 4 (\t) \nno
\mmm
&&
\kern-25pt
+ {{\b\apr}\over 4} \ln\lsqq - {1\over{L\uu{\prime 2}}}
(\s\uu + - \t\uu + )(\s\uu - - \t\uu - ) \rsqq
\int\ll{\s\uu -}\uu{\t\uu -}
dy N(\t\uu +, y) \tilde{b}\ll 4  (y) \ ,
\een{opeycapp}
\bbb
Y\uu - (\s) \tilde{c}\ll 4 (\t) &\simeq&
\nno Y\uu - (\s) \tilde{c}\ll 4 (\t) \nno
\mmm
&&
\kern-25pt
- {{\b\apr}\over 4} \ln\lsqq - {1\over{L\uu{\prime 2}}}
(\s\uu + - \t\uu + )(\s\uu - - \t\uu - ) \rsqq
\int\ll{\s\uu +}\uu{\t\uu +}
dy N(y, \t\uu -) b\ll 4 (y) \ ,
\een{opeyctildeapp}
modulo the addition of smooth terms.

\subsection{Quantum change of variables for composite operators}
\label{method}
Having established corrections to the OPEs
involving the basic fields in both the \ir and \uv
variables, we now present some OPEs involving 
combinations of fields that will be directly relevant 
to the renormalization of the stress tensor.
Given our normal ordering definitions, we can
apply the change of variables in Eqns.~\rr{xshift} and \rr{lorentzinvar0} 
to composite operators.  We thus calculate the quantum corrections
in four steps.  First, we begin by defining normal-ordered composite operators 
in the \uv fields, written as $\s\to\t$ limits 
of unordered products of separated fields, with singularities
subtracted explicitly.  Second, we write the unordered products of \uv fields,
for finite $\s - \t$, in terms of the \ir fields, applying the change
of variables in Eqn.~\rr{lorentzinvar0}.  
The resulting expression is an unordered product
of \ir variables at separated points $\s,~\t$ with a singularity 
subtracted.  Third, we use the OPE of the \ir description to rewrite
the product in terms of normal-ordered $\nno\cc\cc \nno$ products.
Finally, we take the limit $\s\to \t$.


If carried out correctly, the above steps always
take a nonsingular expression in the \uv variables
to a nonsingular expression in the \ir variables.
As an example, we can compute the
transformation of the composite operator
$:\psi\uu + (\s) \pp\ll + \psi\uu - (\s):$.
Carrying out the above steps explicitly, we
rewrite the expression as
\bbb
\lim\ll{\s\to \t} ~\pp\ll{\t\uu +}\lrdd \psi\uu + (\s) \psi\uu - (\t)
- {i\over{\t\uu + - \s\uu +}} \rrdd\ .
\nn
\eee 
For convenience, we can
simply calculate the transformation
of $:\psi\uu + (\s) \psi\uu - (\t) :$ directly, 
differentiating with respect to $\t\uu +$ before taking 
the limit $\s\to \t$.

We next rewrite the combination $\psi\uu + (\s) \psi\uu - (\t) 
- {i\over{\t\uu + - \s\uu +}}$ as
\bbb
- {i\over{\t\uu + - \s\uu +}} +
2 \biggl[ \pp\ll + c\ll 4 (\s) +  \b (\pp\ll + X\uu +)(\s) c\ll 4 (\s) 
\biggr] \lrdd M(\t) \tilde{c}\ll 4  (\t) \rrdd\ .
\nn
\eee 
We now refer to the OPE 
of $c_4(\s)$ with $\tilde {c}_4(\t)$ \rr{opecctilde} and expand 
about small separations $\e\uu \pm  \equiv \s\uu\pm - \t\uu\pm$:
\bbb
c_4(\s) \tilde{c}_4(\t) &=& \nno  c_4(\s) \tilde{c}_4(\t) \nno 
\mmm
&&
- {{i}\over 2} 
\ln \lba  \e\uu + \e\uu - / L\psq \rba \cdot
\lrdd N(\t)  + \e\uu + \pp\ll + N (\t) + \hh \e\uu{+2}
\pp\ll +^2 N (\t) \rrdd
\mmm
&&
+ {{i}\over 2}
\lrdd \e\uu + \pp\ll + N (\t) + {3\over 4} \pp\ll +^2 N (\t)
\e\uu{+2}\rrdd
\mmm
&&
+ {{i}\over 2}\e\uu -
\lrdd - \pp\ll - N(\t)  - \e\uu + \pp\ll + \pp\ll - N (\t) - {1\over 4}
\e\uu {-} \pp\ll -^2 N (\t) \rrdd + O(\e\uu 3)\ . 
\eee
Using this, we rewrite the operator as
\bbb
:\psi\uu + (\s) \psi\uu - (\t): =
2~\nno \lrdd \pp\ll + c\ll 4 (\s) + \b \pp\ll + X\uu + (\s) c\ll 4(\s)
\rrdd\cc\tilde{c}\ll 4 (\t) M(\t)\nno 
\nn\\\nn\\
- i \b (\s\uu + - \t\uu +) \pp\sqd\ll + X\uu + (\t)
+ O((\s- \t)\sqd)\ .
\eee
In particular, we find that the singularity is canceled.
As our final step, we take a derivative with respect to $\t\uu +$ and
find the limit as $\s\to \t$.  We are left with
\bbb
:\psi\uu + (\pp\ll + \psi\uu -) :
= \nno (\pp\ll + c_4 + \b c_4  \pp\ll + X\uu + ) (b_4 + 
2\b M \tilde{c}_4 \pp\ll + X\uu +) \nno + i \b   \pp\ll +^2 X\uu + \ .
\een{quantumcorrected1}
Applying the change of variables in Eqn.~\rr{lorentzinvar0} as a
classical field transformation yields the first term
in Eqn.~\rr{quantumcorrected1}, but not the second.  
We should therefore think of $ \b   \pp\ll +^2 X\uu +$ as a 
quantum correction to the transformation of $: \psi\uu + \pp\ll + \psi\uu -:$.

\subsection{Transformation of the stress tensor and supercurrent} 
Using the algorithm described above, 
we can transform the various terms in the stress tensor, including
quantum corrections.   The quadratic term in the stress
tensor $T\uu{X\uu\m}$ transforms as
\bbb
{2\over{\apr}}\cc:(\pp\ll + X\uu +)(\pp\ll - X\uu -): &=& 
 {2\over{\apr}}\cc\nno (\pp\ll + Y\uu +)(\pp\ll - Y\uu -) \nno 
+ 2 i\b\sqd  M (\pp\ll + X\uu +)\sqd \nno c_4 \tilde{c}_4 \nno
\mmm
&&
+ 2 i \b  M \pp\ll + X\uu + \nno \pp\ll + c_4 \tilde{c}_4 \nno
+ i \b \pp\ll + X\uu + \nno c_4 b_4 \nno\ ,
\eee
with no quantum correction relative to that which is obtained by
classical substitution of Eqn.~\rr{lorentzinvar0} alone.
The linear dilaton term $V\ll\m \pp\ll + \sqd X\uu\m$
transforms straightforwardly, again with no quantum correction:
\bbb
\apr V\ll\m \pp\ll + \sqd X\uu\m &=& V\ll\m \pp\ll + \sqd Y\uu\m  
- i ~ 
\nno 
\Bigl( 
\b\sqd (\pp\ll + Y\uu +)\sqd  M c\ll 4 \tilde{c}\ll 4
+ \b (\pp\ll + \sqd Y\uu +) M c\ll 4 \tilde{c}\ll 4 
\mmm
&&
+ M c\ll 4\prpr \tilde{c}\ll 4 
+ 2\b M (\pp\ll + Y\uu +) (\pp\ll + c\ll 4) \tilde{c}\ll 4 
\mmm
&&
+ \hh c\ll 4  (\pp\ll + b\ll 4) + (\pp\ll + c\ll 4) b\ll 4 + \hh\b (\pp\ll + 
Y\uu +) c \ll 4 b\ll 4 
\Bigr) \nno \ .
\eee
The only terms whose
transformation receives a quantum correction are
the terms in the fermionic stress tensor $T\uu{\psi\uu\m}$
in \rr{stress0}.


The normal-ordered product of the two
$\psi$ fermions is given in Eqn.~\rr{quantumcorrected1}.
Differentiating and taking
the limit $\s\to\t$, we derive the quantum-corrected
transformations for the fermion terms in the stress tensor:
\bbb
- {i\over 2} : \psi\uu - (\pp\ll + \psi\uu +) :
- {i\over 2} : \psi\uu + (\pp\ll + \psi\uu -)  :
&=&  
 - {i\over 2}~
\nno \Bigl(  (\pp\ll + c_4) b_4 + \b \pp\ll + X\uu + c_4 b_4 
\mmm
&&
\kern-180pt
+ 2 \b\sqd  (\pp\ll + X\uu +)\cc M \cc \sqd c_4 \tilde{c}_4
- 2 M (\pp\ll + \sqd c_4) \tilde{c}_4
- 2 \b M \pp\ll +^2 X\uu + c_4\tilde{c}_4 
\Bigr) \nno 
+  \b \pp\ll +^2 X\uu + \ . 
\een{quantumcorrectedfermionstresstensor}
Summing the various remaining terms in the stress tensor, we find
\bbb
\tlc &\equiv & T\uu{X\uu\m} + T\uu{\psi\uu\m}\ ,
\mmm
T\uu{Y\uu\m} + T\uu{\psi\uu\m} & = &  
   - \frac{1}{\a'} G\ll{\m\n}
  \pp_+  Y^\m \pp_+ Y^\n  + V\ll\m \pp\sqd Y\uu\m  
\mmm
&&
	- \frac{3i}{2}  \pp_+ c\ll 4\, b\ll 4 
	-\frac{i}{2} c\ll 4\, \pp_+ b\ll 4 
        +  \b \pp\ll +^2 X\uu +\ .
\een{quantumstress2}

The supercurrent in \uv variables is naturally
decomposed into four pieces, written explicitly in 
Eqn.~\rr{supercurrentpieces} above.
Of the four pieces, only ${\bf 1}$ receives a
quantum correction to its classical transformation
under the change of variables.  To transform ${\bf 1}$ 
with quantum corrections included, we make use of
the OPE
\bbb
(\pp\ll + X\uu - (\s)) \psi\uu + (\t)
&= &
2~ \nno \pp\ll {\s\uu +} \lrdd Y\uu  -(\s) + i \b \apr
M(\s) c_4(\s) \tilde{c}_4(\s) \rrdd 
\mmm
&&
\kern-00pt
\times 
\lrdd \pp\ll + c_4(\s) 
+ \b \pp\ll +Y\uu +(\t) c_4(\t) \rrdd \nno
\mmm
&&
\kern-40pt
+ {{\b\apr}\over 2} \cc \pp\ll +^2 c_4(\s) 
+ \b\sqd \apr c_4(\s) \pp\ll +^2 Y\uu +(\s) 
+ O(\s - \t) + O(M\uu{-1})\ , 
\eee
which we derived according to the rules in Section \ref{method}.
The term ${\bf 1}$ in the supercurrent then transforms
into ${\bf 1}\ll{\rm classical} + {\bf 1}\ll{\rm quantum}$,
where
\bbb
{\bf 1}\ll{\rm classical} &=& 
- 2 \sqrt{2\over\apr}
\nno 
\pp\ll +
\lrdd Y\uu - + i \b\apr M  c_4
\tilde{c}_4  \rrdd  \lrdd \pp\ll + c_4 + \b \pp\ll + Y\uu +  c_4 \rrdd \nno\ ,
\mmm
{\bf 1}\ll{\rm quantum} &=& - \b\sqrt{{{\apr}\over 2}}
 \pp\ll +^2 {c_4}
-2 \b\sqd \sqrt{ {{ \apr}\over 2}}
c\ll 4 \pp\ll +^2 Y\uu + \ .
\een{transformofsupercurrentpieceone}

The remaining three parts are given simply by classical
substitution, because the corresponding operator 
products have no singularities from the interaction terms:
\bbb
{\bf 2} &=& - \sqrt{2\over\apr}
M (\pp\ll + Y\uu +) \tilde{c}_4\ ,
\mmm
{\bf 3} &=& 
2 q \sqrt{\apr} (\pp\ll +\sqd c\ll 4) + 2 \b q \sqrt{\apr} 
(\pp\ll + Y\uu +)(\pp\ll + c\ll 4) + 2 \b q \sqrt{\apr} c\ll 4 (\pp\ll +
\sqd Y\uu +)\ ,
\mmm
{\bf 4}  &=& q \sqrt{\apr} M~(\pp\ll + \tilde{c}\ll 4)
+ q\sqrt{\apr}  \b \tilde{c}_4~ M (\pp\ll + Y\uu +)\ .
\eee
Using fermi statistics, the equations of motion for $\tilde{c}_4$, and
the relation $\b q = {{\sqrt{2}}\over{\apr}}$, we 
recover a form of the supercurrent that is manifestly
finite in the limit $M\to\infty$.  The various terms appear as
\bbb
{\bf 1}\ll{\rm classical} 
&=& - 2 \sqrt{2\over{\apr}}~
\nno \lsqq (\pp\ll + c\ll 4) (\pp\ll + Y\uu -)
+ \b c\ll 4 (\pp\ll + Y\uu +) (\pp\ll + Y\uu -) 
- {{i\b\apr}\over 2}  (\pp\ll + c\ll 4) b\ll 4  c\ll 4   \rsqq \nno\ ,
\mmm
{\bf 1}\ll{\rm quantum} &=& - \b\sqrt{{{\apr}\over 2}}
 \pp\ll +^2 {c_4}
-2 \b\sqd \sqrt{ {{ \apr}\over 2}}
c\ll 4 \pp\ll +^2 Y\uu +\ ,
\mmm
{\bf 2} + {\bf 4} &=&  {q\over 2}  \sqrt{\apr} b\ll 4\ ,
\mmm
{\bf 3} &=& 
2 q \sqrt{\apr} (\pp\ll +\sqd c\ll 4) + 2 \sqrt{2\over\apr}
(\pp\ll + Y\uu +)(\pp\ll + c\ll 4) + 2 \sqrt{2\over\apr} c\ll 4 (\pp\ll +
\sqd Y\uu +)\ .
\ssn{finalsupercurrent1app}


\appendix{BRST cohomology in the Berkovits-Vafa formalism}
\label{BRST}


To show that the Berkovits-Vafa formalism is equivalent to the usual formulation of
the bosonic string, one can use a local operator transformation that
puts the BRST current into a form that makes the equivalence manifest.\footnote{The authors
thank N. Berkovits for explaining this procedure to us.}  We note that
in this appendix we refer to the usual $\b\g$ Fadeev-Popov ghost system using the
conventional notation.  The reader should take care not to confuse the $\b$ ghost with
the lightlike Liouville exponent $\b$, which plays no role in this appendix.  

The BRST current of the Berkovits-Vafa formalism is of the same form as any other
superstring theory in covariant gauge.  
It is convenient to break quantities into matter and reparametrization
ghost sectors.  We use the notation $\bfp$ and $\cfp$ to denote the Fadeev-Popov
$bc$ ghost system.  As noted, the $\b \g$ system will be denoted in the standard way.
The BRST current appears as
\bbb
\jbrst = \cfp \lrdd \tfif + \hh T\uu{\rm gho} \rrdd
+ \g \lrdd G\uu{\rm bv} + \hh G\uu{\rm gho}\rrdd \ ,
\eee
where 
\bbb
T\uu{\rm gho} &\equiv&
- i (\pp\ll + \bfp) \cfp 
- 2 i \bfp (\pp\ll + \cfp) 
- \hh (\pp\ll + \b) \g 
- {3\over 2} \b(\pp\ll + \g) \ ,
\xxx
G\uu{\rm gho} &\equiv& 
(\pp\ll + \b)\cfp 
+ {3\over 2} \b (\pp\ll + \cfp) 
- 2 \bfp\g \ ,
\xxx
\tfif &\equiv& T\uu{\rm mat} + T\uu{b\ll 1 c\ll 1}\ ,
\eee
and $G\uu{\rm bv}$ is the Berkovits-Vafa supercurrent \rr{bvsupercurrent};
here we use the explicit label ``bv'' to distinguish this
supercurrent from that of the ghost (gho) sector.
The notation $\tfif$ is meant to indicate the full stress tensor 
$T\uu{\rm mat} + T\uu{b\ll 1 c\ll 1}$ appropriate for a $c = 15$ SCFT.

The transformation we seek is generated by a Hermitian generator $h$
(here and throughout this subsection, normal ordering is implied
unless otherwise specified):
\bbb
h &\equiv& + {1\over {2\pi}} \int d\s\ll 1 \Bigl[
 \hh \bfp\cfp(\pp\ll +  c\ll 1) c\ll 1 
- i \bfp \g c\ll 1 
- {i \over 4}  \b\g (\pp\ll + c\ll 1) c\ll 1 
\xxx
&&
-i  \cfp \b (\pp\ll + c\ll 1) 
+\frac{i}{2} (\pp\ll + \cfp) \b c\ll 1
\Bigr]\ .
\eee
Defining the unitary operator $U\equiv \exp{i h}$, we can transform the dynamical variables
of the theory into a set of variables in which the BRST current assumes a simple
form.  In particular, the $\b,\g,b\ll 1,c\ll 1$ fields are
decoupled from the remaining degrees of freedom, in transformed variables.  

For completeness, we record the OPEs of the 
$\bfp,\cfp,\b,\g$ and $\bone,\cone$ fields:
\bbb
\bfp(\s) \cfp(0) &\sim  &   -\frac{i}{\s\uu + + i \e } \ , 
\xxx
 \bone(\s) \cone(0) &\sim  &   +\frac{i}{\s\uu + + i \e } \ , 
\xxx
\beta(\s) \gamma(0)&\sim  &   -\frac{1}{\s\uu + + i \e } \ . 
\eee
The transformation of the BRST current of the full theory is thus obtained 
as follows:
\bbb
\jbrst\uu{\new} &\equiv&  U \, \jbrst \, U^{-1}
\xxx
& = &  \jbrst\uu{\bos} + \jbrst\uu{\quartet} +
\jbrst\uu{\deriv} \ ,
\eee
where we have defined
\bbb
\jbrst\uu{\bos}   & \equiv & c T\uu{\bos} - i :b (\pp\ll + c) c:
+ {3\over 2} \pp\ll + \sqd c \ ,
\xxx
\jbrst\uu{\quartet} & \equiv & b\ll 1 \g \ ,
\xxx
\jbrst\uu{\deriv} & \equiv &
\pp_+ \Bigl(  
- {i\over 4} \bfp \cfp \g c\ll 1
- {1\over 4} \b \g\sqd c\ll 1
+ {i\over 2} \cfp b\ll 1 c\ll 1
- {3\over 4} \cfp \b \g
- {i\over 8} \cfp \b \g (\pp\ll + c\ll 1) c\ll 1
\xxx
&  &
 - {{11 i}\over {16}} \cfp (\pp\ll + \sqd c\ll 1) c\ll 1
- {5\over 4} \g(\pp\ll + c\ll 1)
+ {{3 i }\over 8} (\pp\ll + \cfp) \cfp \b c\ll 1
\xxx
&&
- {{5 i }\over 8} (\pp\ll + \cfp) (\pp\ll + c\ll 1) c\ll 1
- \hh (\pp\ll + \cfp)
 \Bigr)  \ .
\eee
The term $\jbrst\uu{\deriv}$ is a total derivative,
and does not contribute to the integrated BRST charge.  The
other two terms integrate to two anticommuting
BRST operators $Q\ll{\rm BRST}\uu{\bos}$ and $Q\ll{\rm BRST}
\uu{\quartet}$.  The BRST cohomology in the `decoupled' variables
must therefore decompose into states of the form $\kket {{\bos}} \otimes
\kket{{\rm quartet}}$, where $\kket{{\bos}}$ and $\kket{{\rm quartet}}$ are in
the cohomology of $Q\ll{\rm BRST}\uu{\bos}$ and $Q\ll{\rm BRST}\uu{\rm quartet}$,
respectively.

The cohomology of $Q\ll{\rm BRST}\uu{\bos}$
is well-known: it consists of operators of the
form $\cfp V\uu{\bos}$ and $\cfp (\pp\ll + \cfp) V\uu{\bos}$,
where $V\uu{\bos}$ is a Virasoro primary of weight one.
The latter, as usual, do not define physical states but rather
the set of `covectors' that have natural inner products 
with the usual physical states \cite{brstcoho}.

The cohomology of $Q\ll{\rm BRST}\uu{\quartet}$ is
straightforward to calculate, since the operator
is exactly quadratic in free fields.  The cohomology
depends only on the `picture' in which
the state is built: that is, the smallest
number $n$ such that $\g\ll{n+{3\over 2}}$ acts
as an annihilation operator and $\b\ll{-(n + {3\over 2})}$ acts as a
creation operator.  So in particular, the entire cohomology
in the 0 picture of the $b\ll 1, c\ll 1, \b,\g$ sector
is exhausted by the identity operator $1$.  The cohomology
in the $-1$ picture is exhausted by the operator $c\ll 1 \d(\g)$.
The cohomology in the $-2$ picture is exhausted by
$c\ll 1 (\pp\ll + c\ll 1) \d(\g) \d(\pp\ll + \g)$.
In each case, the entire cohomology
consists of a single state, which can be represented as the
product of a state in the $\b,\g$ sector with a state in the $b\ll 1,c\ll 1$
sector.  The state of the $\b,\g$ sector is a 
generalized `vacuum', defined as the state annihilated by 
all the annihilation operators $\g\ll{{\rm picture} + m + {3\over 2}},~
\b\ll{m-\hh-{\rm picture}}$ for $m\geq 0$.  The state of the $b\ll 1, c\ll 1$
sector is the state of lowest weight whose value of $n\ll{b\ll 1}- n\ll{c\ll 1}$ is
equal to the picture of the state.

The total ghost number $n\ll {\cfp} + n\ll \g - n\ll{\bfp} -
n\ll \b$ is a grading respected by the BRST charge (which has
ghost number $+1$) and also by our unitary transformation
$U\equiv \exp{i h}$, since $h$ has ghost number zero.  
It is therefore useful to classify the BRST cohomology by
ghost number.  (We count the Berkovits-Vafa fields
$b\ll 1, c\ll 1$ as having ghost number zero.)  For every value $m$ of
the ghost number, there are exactly two kinds of BRST cohomology
classes, each built from a matter vertex operator $V\ll{\bos}$ that
is a primary of weight one with respect to $T\uu{\bos}$.
The first set is of the form
\bbb
\cfp V\ll {\bos} {\cal O}\ll{m-1} \ ,
\eee
while the second set is of the form
\bbb
\cfp(\pp\ll +\cfp) V\ll{\bos} {\cal O}\ll{m-2}\ .
\eee
${\cal O}\ll{m}$ is the picture $m$ vacuum of the $b\ll 1,c\ll 1,\b,\g$ system,
which is proportional to
\bbb
{\cal O}\ll {m<0} \equiv \prod\ll{j=0}\uu{-m-1} (\pp\uu j\ll + c\ll 1) \d(\pp\ll + \uu j \g) \ ,
\eee
for negative $m$, or
\bbb
{\cal O}\ll {m>0} \equiv \prod\ll{j=0}\uu{m-1} (\pp\uu j\ll + b\ll 1) \d(\pp\ll + \uu j \b)\ ,
\eee
for positive $m$, and is proportional to the identity for $m=0$.

Because the picture zero cohomology is just the identity, we know that the
cohomology at picture $-m$ must consist entirely of the $m^{\rm th}$ power
of the {\it inverse} picture changing operator, 
defined by \cite{witteninversepco,yamroninversepco}
\be
Y  \equiv \cfp \d'(\g) \ .
\label{origformofY}
\ee
In decoupled variables, this operator appears as
\be
Y_{\decup} = -c_1 \delta(\g) \ ,
\ee
up to terms that are BRST trivial.
More explicitly, the two are related by
\be
 U \, Y \, U^{-1} &=& \cfp \d'(\g) - c_1 \d(\g) 
	+ \frac{i}{2} (\pp \ll +\cfp) \cfp c_1 \d'' (\g) 
\xxx
	& \equiv & Y\ll{\decup}
	-\frac{i}{2} Q_{\rm BRST}^\decup \left(  \cfp c_1 \d''(\g) \right)  \ ,
\ee
where we have defined $Q_{\rm BRST}^\decup = Q^{\bos}_{\rm BRST} + Q^{\quartet}_{\rm BRST}$.
In the other direction,
\be
U^{-1} Y\ll{\decup} U 
& = & - Y 
- \frac{i}{2}  Q_{\rm BRST}  \lrdd \cfp c_1 \d''(\g) \rrdd
\ .
\ee 

Starting with a vertex operator in the decoupled 
variables in $-1$ picture
\be
\cfp c\ll 1 \d(\gamma) V_{\bos} \ ,
\ee
it is straightforward (in fact, trivial) to 
translate to the usual Berkovits-Vafa framework. 
The generator $h$ commutes with $V\ll{\bos}$, as well
as with any operator of the form $\cfp c\ll 1 f(\g)$, so
the form of the $-1$ picture physical state vertex operator is
unchanged.

The situation is slightly more complicated 
when starting from another picture.
For example, the picture zero vertex operator $\cfp V\ll{\bos}$ does
not commute with the generator $h$, and it has a nontrivial
(inverse) transformation to the standard Berkovits-Vafa theory.
However, we know explicit forms for all physical states and dual physical states in every
picture in the decoupled variables.  For instance, the
physical state and its dual in picture zero are $\cfp V\ll{\bos}$
and $(\pp\ll + \cfp)\cfp V\ll{\bos}$, respectively.   
Explicitly, The inverse transformation defined by the unitary operator $U$ appears as
\bbb
U\uu{-1} \, \cfp V\ll{\bos} \, U & = &  \cfp V\ll{\bos} - \g c\ll 1 V\ll{\bos} \ ,
\xxx
U\uu{-1}\, (\pp\ll + \cfp) \cfp V\ll{\bos} \, U & = &
\Bigl( \cfp (\pp\ll +  c\ll 1) \g + \cfp c\ll 1 (\pp \ll + \g) 
+ \g\sqd (\pp\ll + c\ll 1) c\ll 1 
\xxx
&&
\kern+30pt
+ (\pp\ll + \cfp ) \cfp - (\pp\ll + \cfp) c\ll 1 \g \Bigr) V\ll{\bos}\ .
\eee
The physical states and dual physical states at picture $-1$ transform trivially:
\bbb
U\uu{-1} \, \cfp c\ll 1 \d(\g) V\ll{\bos} \, U & = & \cfp c\ll 1 \d(\g) V\ll{\bos} \ ,
\xxx
U\uu{-1} \, (\pp\ll + \cfp) \cfp c\ll 1 \d(\g) V\ll{\bos} \, U & = &  
(\pp\ll + \cfp) \cfp c\ll 1 \d(\g) V\ll{\bos}\ .
\eee
These objects exhaust the BRST cohomology at pictures 0 and 1 in the Berkovits-Vafa
string.  We will use our knowledge of the full BRST cohomology in the next subsection.

\appendix{Ramond sectors}

The late-time limit of our model differs from the formulation of \cite{bv}
in one respect.  In the usual framework of 
\cite{bv}, all operators in the sector $\tfif$
are neutral under the GSO projection, and 
all fields have their usual periodicity in
the Ramond sectors, except for the $b\ll 1 c\ll 1$
fields. 
The result is that, in \cite{bv}, there are no physical
states in the R/R sector.
The R/R sector has four vacua, but the
one that obeys the physical state conditions
$G\ll 0 = \tilde{G}\ll 0 = 0$ does not
survive the GSO projection.
Modular invariance requires that $\mfw$ act on 
$\kket{\uparrow \ll{\rm L} \otimes \downarrow \ll{\rm R}}$
and $\kket{\downarrow \ll{\rm L} \otimes \uparrow \ll{\rm R}}$
with a $+1$ sign and
 $\kket{\uparrow \ll{\rm L} \otimes \uparrow \ll{\rm R}}$
and $\kket{\downarrow \ll{\rm L} \otimes \downarrow \ll{\rm R}}$
with a $-1$.
Here, $\kket{\downarrow\ll{\rm R}}$ is defined to be
annihilated by the zero mode $b_1^0$, and similarly for $\kket{\downarrow\ll{\rm L}}$
and $\tilde b_1^0$.
The physical state conditions in the R/R sector are that
the matter must be primary of weight $1$, and that the \zombies lie
in the state $\kket{\downarrow \ll{\rm L} \otimes \downarrow \ll{\rm R}}$.
The primary state in the $b_1 c_1 \tilde b_1 \tilde c_1$ sector 
violates the GSO projection.  In the original Berkovits-Vafa 
embedding \cite{bv}, the matter in $T^{\rm mat}$ is all GSO neutral 
so there are no physical states surviving the GSO projection in the R/R sector.  

By contrast, the R/R sector in our theory is nonempty, because
the operator $\mfw$ acts not only on $b\ll 1 c\ll 1
\tilde{b}\ll 1 \tilde{c}\ll 1$, but also
on the $\psi\uu{Z\uu i}$ and $\pst\uu{Z\uu i}$ fields.
All worldsheet fermions are periodic in the R/R sectors,
so it is possible to change the
effective GSO projection in the $b\ll 1 c\ll 1\tilde{b}\ll 1
\tilde{c}\ll 1$ sector by acting with a zero mode of one
of the $\psi\uu{Z\uu i}$ or $\pst\uu{Z\uu i}$ states.
The spectrum of physical states in the R/R sector is
therefore nonempty and contains bispinor states of
the $SO(D-2)\ll{\rm L} \otimes SO(D - 2)\ll{\rm R}$ current
algebra describing the dynamics of the $\psi\uu{Z\uu i},~ \pst\uu{Z\uu i}$ fields.
The p-form R/R fluxes of the original type 0 theory
turn into these.


Let us emphasize that the presence of a nontrivial R/R sector in this theory
does not imply the existence of any {\it additional} states in the theory, relative
to the original Berkovits-Vafa embedding described in \cite{bv}.  
If one were to omit the R/R sector altogether, this would
leave a tree-level spectrum equivalent to a subsector of a consistent bosonic string
spectrum that, by itself, is not modular-invariant.  The states appearing in the NS sector
are all vector and tensor sates of the current algebra group $SO(D-2)_L \otimes SO(D-2)_R$,
and give rise to a partition function that is not invariant under $ \tau \to -1/\tau$.
The bispinor states appearing in the R/R sector restore modular invariance, and are thus 
necessary for the consistency of the theory.  They do {\it not} add any additional
sectors relative to the states of bosonic string theory.  

In other words, the standard formulation of the bosonic string has
states of the form
\bbb
\kket{\rm tensor} ,~~\kket{\rm spinor} ,
\eee
which are represented as vertex operators of the form
\bbb
V\uu{\bos}\ll{\rm tensor}\ , ~~V\uu{\bos}\ll{\rm spinor}\ ,
\eee
where $V\uu{\bos}$ is a local operator of weight 1, constructed from the
current algebra and $X\uu\m$ degrees of freedom
.
Modular invariance demands that both types of operators are included.

In the Berkovits-Vafa embedding that emerges naturally in our model, these states appear 
as vertex operators of the form
\bbb
\cfp\,  c\ll 1 \, \d(\gamma) \, V\uu{\bos}\ll{\rm tensor}\,
	~~~{\rm and}~~~\cfp\, \Sigma\uu{b\ll 1 c\ll 1} \,
\exp{- \phi/2} \, V\uu{\bos}\ll{\rm spinor} \, 
\eee
respectively, where $\Sigma\uu{b\ll 1 c\ll 1}$ is the ground state spin field for the
$b\ll 1, c\ll 1$ system (i.e.,~the lowest weight operator that creates a $\IZ\ll 2$ branch cut
in the conjugate pair $b\ll 1, c\ll 1$).  The field $\phi$ is the chiral boson that appears in the
usual bosonization of the $\b,\g$ superghost system, with the identification $\delta(\gamma)
= \exp{-\phi}$.  Both types of states are BRST invariant
and exhibit properties identical to
the corresponding bosonic string states.


As noted, 
our embedding admits Ramond sectors that realize spinor states of
the current algebra, while the spectrum of the Berkovits-Vafa model presented in
\cite{bv} is of the form
\bbb
V\uu{\bos}\ll{\rm tensor}\, c\ll 1\, \d(\g),~~~
	{\rm and}~~~V\uu{\bos}\ll{\rm spinor}\, c\ll 1\, \d(\g)  \ ,
\eee
with all physical vertex operators realized as NS states.

All three possible presentations of the spinor states are equivalent at the level of the spectrum:
the combinations $\Sigma\uu{b\ll 1 c\ll 1}\exp{-\phi/2}$
in our embedding and $c\ll 1 \d(\g)$ in the
embedding of \cite{bv} both have weight zero and function only to dress a matter vertex operator
in a BRST-invariant way.  Since the only role of the $b\ll 1, c\ll 1$ system is to ``offset'' or ``cancel'' 
the contribution of the superghosts $\b,\g$, it should not matter whether the $b\ll 1, c\ll 1$
and $\b,\g$ systems are simultaneously periodic or simultaneously antiperiodic, so long as the
dressed vertex operators are BRST-invariant.


Given this argument, it should be expected that the physical states, as presented above,
should exhibit the same interactions as those appearing in the standard Berkovits-Vafa
system.  Is suffices to show that the operator products of physical state vertex operators
are independent of the presentation.  We can show this explicitly by considering the OPE of
two fixed-picture vertex operators in the Berkovits-Vafa formalism of \cite{bv}.
We will label the bosonic string vertex operators as $V^{\bos}_A$ and $V^{\bos}_B$,
focusing only the right-moving sector; the corresponding arguments in
the left-moving sector follow by analogy.

Both bosonic vertex operators are Virasoro primaries of weight 1 in the $c=26$ matter system
$T\uu{{\bos}}$.  Their OPE is of the form
\be
V^{\bos}_A (\s) ~V^{\bos}_B (0) = \sum\ll D
f\ll{AB}{}\uu D (\s\uu + + i \e)\uu{\Delta\ll D - 2} {\cal O}\uu{\bos}\ll D\ ,
\label{theope}
\ee
where ${\cal O}\uu{\bos}\ll D$ is a third operator of weight $\Delta\ll D$.
Only primary operators with $\Delta\ll D = 1$ can appear in BRST-nontrivial terms on the right-hand
side.  We focus on these and label them $V\uu {\bos}\ll D$.

The $-1$ picture vertex operator is of the form
\be
\cfp\, \delta(\gamma)\, c\ll 1\, V\uu{\bos}\ll D   \  ,
\ee
and the $0$ picture vertex operator appears as
\be
\cfp\, V^{\bos} - \gamma\, c_1\, V^{\bos}\ll D  \ .
\ee
Since both operators on the left-hand side are BRST
closed, all operators appearing on the right-hand side
will be BRST closed as well.
As explained in Ref.~\cite{bv} and above,
the only BRST nontrivial terms on the right-hand side are
those with dressing $c_1 \delta(\gamma)$ and a primary of weight one in the bosonic
sector.  So we discard all operators on the right-hand side except those whose $b\ll 1 c\ll 1$
content is $c\ll 1$, whose $\b\g$ content is $\delta(\gamma)$ and whose bosonic
content is $V\uu{\bos}\ll D$, with conformal weight $1$;  
the discarded terms are all BRST-trivial.
It follows that, up to BRST equivalence, the
product of the two operators above is given by
\be
(\pp\ll + \cfp)\, \cfp \, c_1\,  \delta(\gamma)\, V_D^{\bos} + O(\s\uu + )  \ .
\ee
Both vertex operators are translationally invariant in BRST cohomology, so we can
discard all terms of $O(\s\uu +  )$ and higher.

Now, suppose that the two operators on the left-hand side are spinor states of the
current algebra group $SO(D-2)_R$.  To demonstrate the physical equivalence
between the presentation at hand and that described in \cite{bv}, we need show only that the
operator products are equal if we replace the spinor state dressings $c_1\, \delta(\gamma)$ with
$ e^{-\phi/2}\, \Sigma^{b_1 c_1}$.
To accomplish this, it is helpful to bosonize the $b_1 c_1$ system in terms of a 
real boson $\htil$ with timelike signature, with OPE
\be
\htil (\s) \htil(0) =  \log (- i (\s\uu + + i \e) )\, + : \htil(\s) \htil (0) :  \ .
\label{HHOPE}
\ee
The stress tensor of the $b_1 c_1$ system in bosonic variables appears as
\be
T = \frac{1}{2} : \pp\ll + \htil \pp\ll + \htil : - \pp\ll +^2 \htil \ .
\ee
The normal-ordered exponential $e^{\Gamma \htil}$ has weight $\frac{1}{2}\G^2 + \G$,
and we bosonize the $b_1 c_1$ system as
\be
c_1 = :e^{-\htil}: \ , \qquad b_1 = :e^{+ \htil}: \ .
\ee
The ground-state spin field $\Sigma^{b_1c_1}$ for the $b_1c_1$ 
system is $e^{-\htil/2}$, with weight $-3/8$.  The OPE of two spin fields is 
\be
\Sigma^{b_1c_2} (\s)\,  \Sigma^{b_1c_2} (0) &=& : e^{-\htil(\s)/2} :\, : e^{-\htil(0)/2} :   
\xxx
	&=&  
e^{-{{\pi i}\over{8}}}(\s\uu + + i \e)^{1/4} : e^{-\htil(\s)/2} e^{-\htil(0)/2} :  \ .
\ee
Expanding around $\s = 0$, we obtain
\be
\Sigma^{b_1c_2} (\s) \, \Sigma^{b_1c_2} (0) &=&  
e^{-{{\pi i}\over 8}} (\s\uu + + i \e)^{1/4} :e^{-\htil(0)}:
\xxx
&&
- \frac{1}{2} e^{-{{5\pi i}\over 8}} (\s\uu + + i \e)^{5/4} 
: \pp\ll + \htil(0) \,  e^{ -\htil(0)} : + O(\s^{9/4}) 
\xxx
\kern-30pt &=&
  e^{- {{\pi i}\over 8}} (\s\uu + + i \e)^{1/4} c_1 
	+ \frac{1}{2} e^{- {{5\pi i}\over 8}} (\s\uu + + i \e)^{5/4} \pp\ll 
	+ c_1 + O(\s^{9/4})  \ .
\ee

We now wish to combine this with the 
superghosts and the bosonic matter components of the theory.
The $-1/2$ picture Ramond vertex operators have ghost dressing $e^{-\phi/2}$.
The OPE analogous to \rr{HHOPE} above appears as
\be
\phi (\s) \phi(0) =  \log ( \s\uu + + i \e )\, + : \phi(\s) \phi (0) :  \ .
\ee
The OPE of two operators $e^{-\phi/2}$ is then given by
\be
: e^{-\phi(\s)/2} :\, : e^{-\phi(0)/2}:  
 	&=& e^{{{\pi i}\over 8}} (\s\uu + + i \e)^{-1/4} \delta(\gamma) 
\xxx
&&	+ \frac{1}{2} e^{- {{3\pi i}\over 8}} (\s\uu + + i \e)^{3/4} (\pp\ll + \gamma) 
		\delta'(\gamma) + O(\s^{7/4}) \ .
\ee
Only the leading terms in the $\Sigma^{b_1c_1} \cdot \Sigma^{b_1c_1}$
and the $e^{-\phi/2} \cdot e^{-\phi/2}$ OPEs will ever contribute nontrivially
to the BRST cohomology, as noted above.
Therefore, the physical component of the operator product in Eqn.~\rr{theope} is
\be
(\pp\ll + \cfp)\, \cfp \, c_1\,  \delta(\gamma)\, V_D^{\bos} + O(\s)  \ .
\ee
Again, the $O(\s)$ and higher terms can be discarded as BRST trivial, since both operators
on the left-hand side of the operator product are translationally invariant in 
BRST cohomology.  
The ring structure of the
BRST cohomology is therefore unchanged under the overall replacement
\be
c_1 \, \delta(\gamma) V^{\bos}_{\rm spinor} \to 
	\Sigma^{b_1 c_1} e^{-\phi/2} V^{\bos}_{\rm spinor} \ .
\ee
We have now established the equivalence of our presentation
of the Berkovits-Vafa system with the original presentation in Ref.~\cite{bv}.

\bibliographystyle{utcaps}
\bibliography{dimchange}

\providecommand{\href}[2]{#2}\begingroup\raggedright\begin{thebibliography}{10}

\bibitem{previous}
S.~Hellerman and I.~Swanson, ``Cosmological solutions of supercritical string
  theory,''
\href{http://www.arXiv.org/abs/hep-th/0611317}{{\tt hep-th/0611317}}.

\bibitem{previous2}
S.~Hellerman and I.~Swanson, ``Dimension-changing exact solutions of string
  theory,''
\href{http://www.arXiv.org/abs/hep-th/0612051}{{\tt hep-th/0612051}}.

\bibitem{bv}
N.~Berkovits and C.~Vafa, ``On the Uniqueness of string theory,'' {\em Mod.
  Phys. Lett.} {\bf A9} (1994) 653--664,
\href{http://www.arXiv.org/abs/hep-th/9310170}{{\tt hep-th/9310170}}.

\bibitem{va}
D.~V. Volkov and V.~P. Akulov, ``Possible universal neutrino interaction,''
  {\em JETP Lett.} {\bf 16} (1972)
438--440.

\bibitem{evaofer}
O.~Aharony and E.~Silverstein, ``Supercritical stability, transitions and
  (pseudo)tachyons,'' {\em Phys. Rev.} {\bf D75} (2007) 046003,
\href{http://www.arXiv.org/abs/hep-th/0612031}{{\tt hep-th/0612031}}.

\bibitem{hellerman2}
S.~Hellerman and X.~Liu, ``Dynamical dimension change in supercritical string
  theory,''
\href{http://www.arXiv.org/abs/hep-th/0409071}{{\tt hep-th/0409071}}.

\bibitem{freedman}
D.~Z. Freedman, M.~Headrick, and A.~Lawrence, ``On closed string tachyon
  dynamics,'' {\em Phys. Rev.} {\bf D73} (2006) 066015,
\href{http://www.arXiv.org/abs/hep-th/0510126}{{\tt hep-th/0510126}}.

\bibitem{seibergcircles}
N.~Seiberg, ``Observations on the moduli space of two dimensional string
  theory,'' {\em JHEP} {\bf 03} (2005) 010,
\href{http://www.arXiv.org/abs/hep-th/0502156}{{\tt hep-th/0502156}}.

\bibitem{tdualitybetweentypeiiandtype0}
O.~Bergman and M.~R. Gaberdiel, ``Dualities of type 0 strings,'' {\em JHEP}
  {\bf 07} (1999) 022,
\href{http://www.arXiv.org/abs/hep-th/9906055}{{\tt hep-th/9906055}}.

\bibitem{mm1}
J.~L. Karczmarek and A.~Strominger, ``Matrix cosmology,'' {\em JHEP} {\bf 04}
  (2004) 055,
\href{http://www.arXiv.org/abs/hep-th/0309138}{{\tt hep-th/0309138}}.

\bibitem{mm2}
J.~L. Karczmarek and A.~Strominger, ``Closed string tachyon condensation at c =
  1,'' {\em JHEP} {\bf 05} (2004) 062,
\href{http://www.arXiv.org/abs/hep-th/0403169}{{\tt hep-th/0403169}}.

\bibitem{mm3}
J.~L. Karczmarek, A.~Maloney, and A.~Strominger, ``Hartle-Hawking vacuum for c
  = 1 tachyon condensation,'' {\em JHEP} {\bf 12} (2004) 027,
\href{http://www.arXiv.org/abs/hep-th/0405092}{{\tt hep-th/0405092}}.

\bibitem{mm4}
T.~Takayanagi, ``Matrix model and time-like linear dilaton matter,'' {\em JHEP}
  {\bf 12} (2004) 071,
\href{http://www.arXiv.org/abs/hep-th/0411019}{{\tt hep-th/0411019}}.

\bibitem{joelectures}
J.~Polchinski, ``What is string theory?,''
\href{http://www.arXiv.org/abs/hep-th/9411028}{{\tt hep-th/9411028}}.

\bibitem{Ohta:1994qp}
N.~Ohta and J.~L. Petersen, ``N=1 from N=2 superstrings,'' {\em Phys. Lett.}
  {\bf B325} (1994) 67--70,
\href{http://www.arXiv.org/abs/hep-th/9312187}{{\tt hep-th/9312187}}.

\bibitem{Bastianelli:1994cb}
F.~Bastianelli, N.~Ohta, and J.~L. Petersen, ``Toward the universal theory of
  strings,'' {\em Phys. Lett.} {\bf B327} (1994) 35--39,
\href{http://www.arXiv.org/abs/hep-th/9402042}{{\tt hep-th/9402042}}.

\bibitem{Bastianelli:1994xb}
F.~Bastianelli, N.~Ohta, and J.~L. Petersen, ``A Hierarchy of superstrings,''
  {\em Phys. Rev. Lett.} {\bf 73} (1994) 1199--1202,
\href{http://www.arXiv.org/abs/hep-th/9403150}{{\tt hep-th/9403150}}.

\bibitem{goldstein}
H.~Goldstein, {\em Classical Mechanics}.
\newblock Addison-Wesley, Reading, MA, second~ed., 1980.

\bibitem{brstcoho}
M.~D. Freeman and D.~I. Olive, ``{BRS cohomology in string theory and the no
  ghost theorem},'' {\em Phys. Lett.} {\bf B175} (1986)
151.

\bibitem{witteninversepco}
E.~Witten, ``{Interacting Field Theory of Open Superstrings},'' {\em Nucl.
  Phys.} {\bf B276} (1986)
291.

\bibitem{yamroninversepco}
J.~P. Yamron, ``{The free Ramond sector of Witten's covariant superstring
  action},'' {\em Phys. Lett.} {\bf B187} (1987)
67.

\end{thebibliography}\endgroup

\end{document}